\documentclass[aps,preprint,showpacs,floats,epsf,epsfig,nofootinbib,12pt]{revtex4}
\usepackage{graphicx}% Include figure files
\usepackage{epsfig}
\usepackage[dvips]{color}
\usepackage{subfigure}

\def\beq{\begin{eqnarray}}
\def\eeq{\end{eqnarray}}

\begin{document}

\title{Study on the possible molecular state composed of $D^*_s\bar D_{s1} $  within the Bethe-Salpeter framework }

\vspace{1cm}

\author{ Hong-Wei Ke$^{1}$   \footnote{Corresponding author, khw020056@tju.edu.cn}, Xiao-Hai Liu$^{1}$ \footnote{xiaohai.liu@tju.edu.cn}and
        Xue-Qian Li$^2$\footnote{lixq@nankai.edu.cn},
   }

\affiliation{  $^{1}$ School of Science, Tianjin University,
Tianjin 300072, China
\\
  $^{2}$ School of Physics, Nankai University, Tianjin 300071, China }

\vspace{12cm}

\begin{abstract}

Recently a vector charmonium-like state $Y(4626)$ was observed in
the portal of $D^+_sD_{s1}(2536)^-$. It intrigues an active
discussion on the structure of the resonance because it has
obvious  significance for gaining a better understanding on its
hadronic  structure with suitable inner constituents. It indeed
concerns the general theoretical framework about possible
structures of exotic states. Since the mass of $Y(4626)$ is
slightly above the production threshold of $D^+_s\bar
D_{s1}(2536)^-$ whereas below that of $D^*_s\bar D_{s1}(2536)$
with the same quark contents as that of $D^+_s\bar
D_{s1}(2536)^-$,  it is natural to conjecture $Y(4626)$ to be a
molecular state of $D^{*}_s\bar D_{s1}(2536)$, as suggested in
literature. Confirming or negating this allegation would shed
light on the goal we concern. We calculate the mass spectrum of a
system composed of a vector meson and an axial vector i.e.
$D^*_s\bar D_{s1}(2536)$ within the framework of the
Bethe-Salpeter equations. Our numerical results show that the
dimensionless parameter $\lambda$ in the form factor which is
phenomenologically introduced to every vertex, is far beyond the
reasonable range for inducing an even very small binding energy
$\Delta E$. It implies that the $D^*_s\bar D_{s1}(2536)$ system
cannot exist in the nature as a hadronic molecule in this model,
so that we may not think the resonance $Y(4626)$ to be a bound
state of $D^*_s\bar D_{s1}(2536)$, but something else, for example
a tetraquark and etc.


\end{abstract}

\maketitle

\section{Introduction}

In 2019 the Belle Collaboration observed a vector charmonium-like
state $Y(4626)$ in the portal of $e^+e^- \to
D^+_sD_{s1}(2536)^-+c.c.$ and its mass and width are
$4625.9^{+6.2}_{-6.0}({\rm stat.} )\pm0.4({\rm syst.} )$ MeV and
$49.8^{+13.9}_{-11.5}({\rm stat.} )\pm4.0({\rm syst.} )$
MeV\cite{Jia:2019gfe}. In 2008 the Belle Collaboration reported a
near-threshold enhancement in the $e^+e^-\to
\Lambda_c^+\Lambda_c^-$ cross section and the peak corresponds to
a hadronic resonance which is named as
$Y(4630)$\cite{Pakhlova:2008vn}. Recently a simultaneous fit was
performed to the data analysis of $e^+e^-\to
\Lambda_c^+\Lambda_c^-$ and a peak with mass and width being
$4636.1^{+9.8}_{-7.2}({\rm stat.} )\pm8.0({\rm syst.} )$ MeV and
$34.5^{+21.0}_{-16.2}({\rm stat.} )\pm5.6({\rm syst.} )$
MeV\cite{Xie:2020dkm} emerges. Due to their very close masses and
widths, it is tempted to consider $Y(4626)$ and $Y(4630)$ are the
same resonance. In Ref.\cite{Wang:2020prx} the authors explained
$Y(4626)$ and $Y(4660)$\footnote{In the 2018 PDG\cite{PDG18}
Y(4660) was named as $\psi(4660)$ and Y(4630) was accounted as the
same meson due to measurement errors. Thus both Y(4660) and
Y(4630) are listed in the PDG  under the $\psi(4660)$ entry.
However there are still diverse views about the structures of
$Y(4660)$ and $Y(4630)$ in the community, in our present work,  we
still use their initial names assigned by the experimentalists who
observed those mesons. We do not suppose they are the same
particle.}\cite{Wang:2007ea} to be mixtures of two excited
charmonia. It also maybe is a non-resonant threshold enhancement due to the opening of the $\Lambda_c^+\Lambda_c^-$ channel
as discussed in\cite{vanBeveren:2008rt,vanBeveren:2010jz}, whereas the
authors\cite{He:2019csk} suggested $Y(4626)$ as a molecular state
$D^*_s\bar D_{s1}(2536)$. In Ref.\cite{Deng:2019dbg,Tan:2019knr}
$Y(4626)$ was regarded as a tetraquark $cs\bar c \bar s$.

Since 2003 many exotic resonances $X$, $Y$ and $Z$
bosons\cite{Choi:2003ue,Abe:2007jn,Choi:2005,Choi:2007wga,Aubert:2005rm,
Ablikim:2013emm,Ablikim:2013wzq,Ablikim:2013mio,Liu:2013dau,Collaboration:2011gja}
have been experimentally observed, such as $X(3872)$, $X(3940)$,
$Y(3940)$, $Z(4430)$, $Y(4260)$, $Z_c$(4020), $Z_c$(3900),
$Z_b(10610)$ and $Z_b(10650)$ (of course, not a complete list). The states
have attracted attention
of theorists because their structures obviously are beyond the simple
$q\bar q$ settings for mesons. If we can firmly determine their
compositions, it would definitely enrich our knowledge on hadron
structures and moreover shed light on the non-perturbative QCD
effects at lower energy ranges.  Studies with different
explanations on the inner structures\cite{Chen:2016spr} have been tried, such as
molecular state, tetraquark or dynamical effect\cite{Guo:2019twa}.
Anyway, all the ansatz have a certain reasonability, but a unique
picture or criterion  for firmly determining the inner structures
is still lacking. Nowadays, the majority of phenomenological researchers
conjectures what the
concerned exotic states are made of, just based
on the available experimental data. Then by comparing the results
with new data one can check the validity degree of the proposal.
If the results obviously contradict to the new measurements of
better accuracy, the ansatz should be abandoned.   Following
this principle we explore $Y(4626)$ by assuming it to be a
molecular state of $D^*_s\bar D_{s1}(2536)$, and then using more
reliable theoretical framework to check the scenario and see if
the proposal from our intuition can be valid.

Concretely, in this work supposing $Y(4626)$ as a $D^*_s\bar
D_{s1}(2536)$ molecular state, we employ the Bethe-Salpeter (B-S)
equation which is a relativistic equation established on the basis
of quantum field theory, to study the two-body bound state
\cite{Salpeter:1952ib}. Initially, the B-S equation was used to
study the bound state of two
fermions\cite{Chang:2004im,Chang:2005sd,Yu:2006ty}, later the
method was generalized to the system of
one-fermion-one-boson\cite{Guo:1998ef}. In
Ref.\cite{Guo:2007mm,Feng:2011zzb} the authors employed the
Bethe-Salpeter equation to study  some possible molecular states,
such as $K\bar K$ and $B\bar K$ system. With the same approach the
bound state of $B\pi$, $D^{(*)}D^{(*)}$, $B^{(*)}B^{(*)}$ are
studied \cite{Ke:2018jql,Ke:2012gm}. Recently the approach was
applied to explore doubly charmed
baryons\cite{Weng:2010rb,Li:2019ekr} and
pentaquarks\cite{Wang:2019krq,Ke:2019bkf}. In this work, we try to
calculate the spectrum of $Y(4626)$ composed of a vector meson and
an axial vector meson.

If two constituents can form a bound state the interaction between
them should be large enough to hold them into a bound state. The
chiral perturbation theory tells us that two hadrons interact via
exchanging a certain mediate meson(s) and the forms of the
effective vertices are determined by relevant symmetries, but the
coupling constants generally are obtained by fitting data. For the
molecular states, since two constituents are color-singlet hadrons
the exchanged particles are some light mesons with proper quantum
numbers. It is noted that even though there are many possible
light mesons contributing to the effective interaction between the
two constituents, generally one or several of them would provide
the dominant contributions. Moreover, beyond it, most of time the
scenario with some other mesons exchange should also be taken into
account, because even though the extra contributions are small
comparing to the dominant one(s), they sometimes are not
negligible, namely it would make the secondary contribution to the
effective interaction. Then the effective kernel for the B-S
equation is set. For the $D^*_s\bar D_{s1}(2536)$ system, the
contribution of $\eta$
\cite{Colangelo:2005gb,Colangelo:2012xi,Ding:2008gr} dominates,
whereas in Ref. \cite{Ding:2008gr} the authors suggested $\sigma$
exchange makes the secondary  contribution.  In our case
considering the concerned quark contents of $D^*_s$ and $\bar
D_{s1}(2536)$,  the contribution of {$\eta'$, $f_0(980)$ and
$\phi$} should stand as the secondary one. The effective
interactions induced by exchanging {$\eta$, $\eta'$, $f_0(980)$
and $\phi$} are deduced with the heavy quark
symmetry\cite{Colangelo:2005gb,Colangelo:2012xi,Ding:2008gr,Casalbuoni:1996pg,Casalbuoni:1992gi,Casalbuoni:1992dx} and
we present the formulas in the appendix A. With the effective
interactions we can derive the kernel and establish the
corresponding B-S equation.

With all necessary parameters being beforehand chosen and input,
the B-S equation is solved
numerically. In some cases the equation does not possess a
solution if one or several parameters are set within a reasonable
range, then
a conclusion must be drawn that the proposed bound state should not exist
in nature. On the contraries, a solution of the B-S equation with
reasonable parameters implies that the corresponding bound state is formed.
In that case, simultaneously the B-S wave function is obtained which can be
used to calculate the rates of
strong decays, which will help
experimentalist to design new experiments for further measurements.

This paper is organized as follows: after this introduction we
will derive the B-S equation related to  possible bound state
composed of $D^*_s$ and $\bar D_{s1}(2536)$ which are a vector and
an axial vector meson respectively. In section III the formula for
its strong decays are present. Then in section IV we will solve
the B-S equation numerically. Since $Y(4626)$ is supposed to be a
molecular bound state, the  input parameters must be within a
reasonable range, but our results say that this mandatory
condition cannot be satisfied, thus we think that such a
molecular state of $D^*_s\bar D_{s1}(2536)$ may not exist.
However, as we deliberately set the parameters to a region which
is not favored by all previous phenomenological works, we can
obtain the required spectrum and corresponding wavefunctions. With
the wavefunction we evaluate the strong decay rate of $Y(4626)$
and present our results by figures and tables. Section IV is
devoted to a brief summary.

\section{The bound states of $D^*_s\bar D_{s1}$}

%\begin{figure}
%\begin{center}
%\begin{tabular}{ccc}
%\scalebox{0.6}{\includegraphics{4625.eps}}\,\,\,\,\,\,\,\,\,\,\,\,\,\,\,\,\,\,\,\,\,\,\,\,\,\,\,\,\,\,\,
%\,\,\,\scalebox{0.6}{\includegraphics{4625p.eps}}
%\end{tabular}
%\end{center}
%\caption{the bound states of  $D^*_s D_{s1}$ formed by exchanging
%$\eta$ }\label{t1}
%\end{figure}

\begin{figure*}
        \centering
        \subfigure[~]{
          \includegraphics[width=7cm]{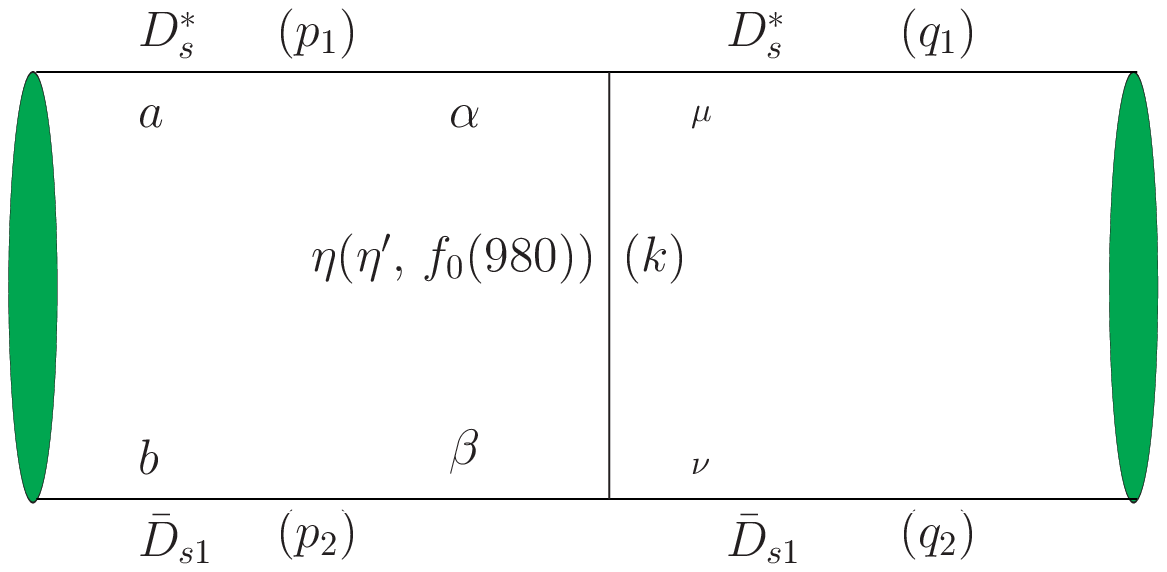}}\,\,\,\,\,\,\,\,\,\,\,\,\,\,\,\,\,\,\,\,\,\,
        \subfigure[~]{
          \includegraphics[width=7cm]{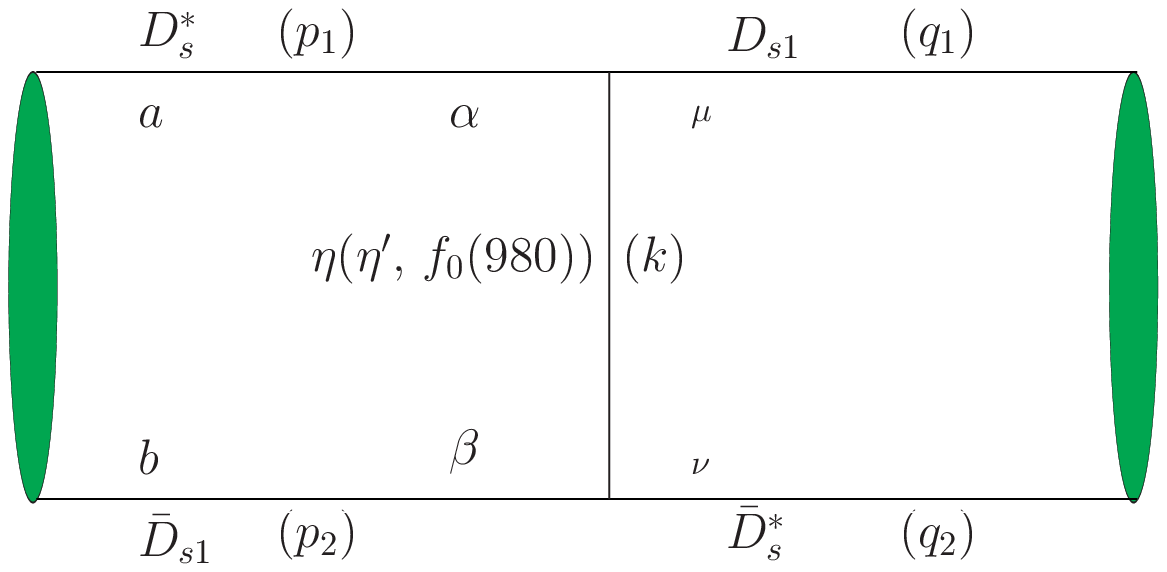}}
 \caption{the bound states of  $D^*_s \bar D_{s1}$ formed by exchanging
$\eta\,(\eta')\,f_0(980)$.}
        \label{t1a}
    \end{figure*}

\begin{figure*}
        \centering
        \subfigure[~]{
          \includegraphics[width=7cm]{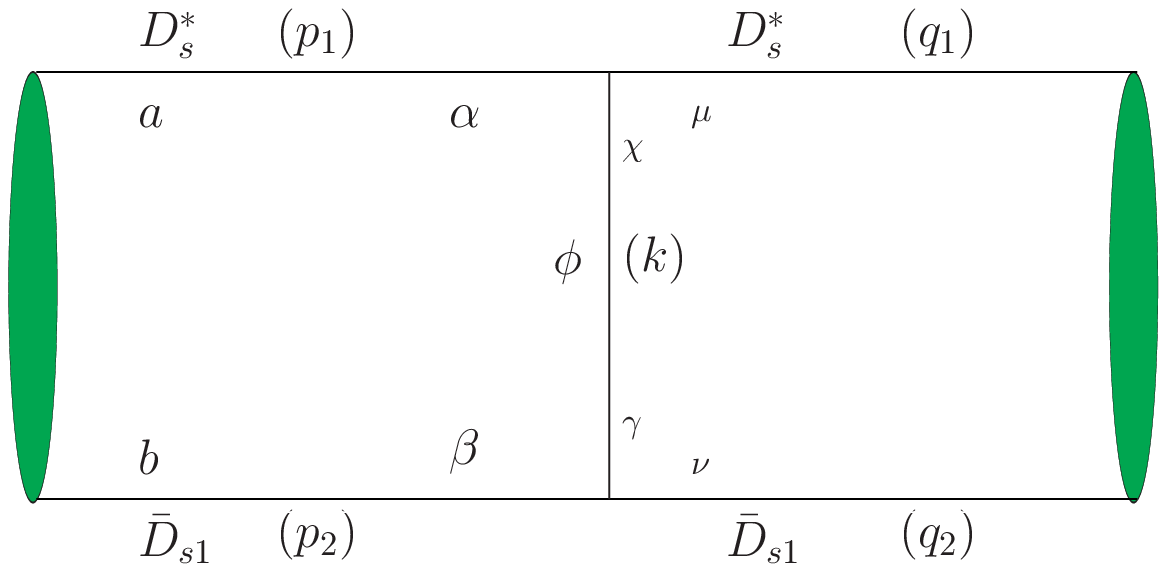}}\,\,\,\,\,\,\,\,\,\,\,\,\,\,\,\,\,\,\,\,\,\,
        \subfigure[~]{
          \includegraphics[width=7cm]{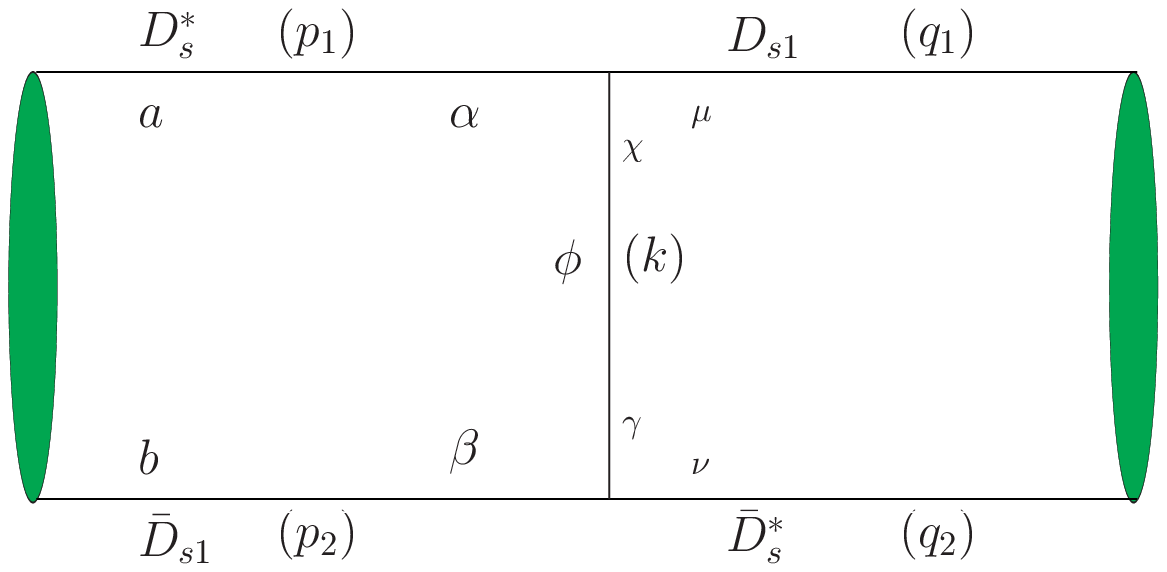}}
 \caption{the bound states of  $D^*_s \bar D_{s1}$ formed by exchanging
$\phi(1020)$.}
        \label{t1b}
    \end{figure*}

Since the newly observed resonance $Y(4626)$ contains hidden
charms and its mass is close to the sum of the masses of $ D^*_s$
and $\bar D_{s1}$ where $ D^*_s-\bar D_{s1}$ corresponds to
$D^{*+}_s-D_{s1}^-$ or $D^{*-}_s-D_{s1}^+$, a
conjecture about its molecular structures composed of $ D^*_s$ and
$\bar D_{s1}$ is favored. For a state with spin-parity being
$1^-$, its spatial wave function is in $S$ wave. There are two
possible states $Y_1=\frac{1}{\sqrt{2}}( D^{*+}_s D_{s1}^-+
D^{*-}_s D_{s1}^+)$ and $Y_2=\frac{1}{\sqrt{2}}( D^{*+}_s
D_{s1}^--D^{*-}_s D_{s1}^+)$. We will focus on such an ansatz and
try to find numerical results by solving the relevant B-S
equation.

\subsection{The Bethe-Salpeter (B-S) equation for $1^-$ $ D^*_s\bar D_{s1}$ molecular state}

By the effective theory  $D^*_s$ and  $\bar D_{s1}$ interact
mainly via exchanging $\eta$. The Feynman diagram at the leading
order is depicted in Fig. \ref{t1a}. To take into account the
secondary contribution induced by exchanging other mediate mesons,
in Ref.\cite{Ding:2008gr} the authors consider a contribution of
exchanging $\sigma$ to the effective interaction. Since there are
neither $u$ nor $d$ constituents in $D^*_s$ and $\bar D_{s1}$,
their coupling to $\sigma$ would be very weak, thus the secondary
contribution to the interaction may come from exchanging
$f_0(980)$ instead. The relevant Feynman diagrams are shown in
Fig. \ref{t1a}. In this work, the contributions induced by
exchanging $\eta'$ (Fig. \ref{t1a}) and $\phi(1020)$ (Fig.
\ref{t1b}) are also taken into account.  The relations between
relative and total momenta of the bound state are defined as
\begin{eqnarray} p = \eta_2p_1 -
\eta_1p_2\,,\quad q = \eta_2q_1 - \eta_1q_2\,,\quad P = p_1 + p_2
=q_1 + q_2 \,, \label{momentum-transform1}
\end{eqnarray}
where $p_1$ and $p_2$ ($q_1$ and $q_2$) are the momenta of the
constituents, $p$ and $q$ are the relative momenta between the two constituents of the bound
state at the both sides of the diagram, $P$ is the total momentum
of the resonance, $\eta_i = m_i/(m_1+m_2)$ and $m_i\, (i=1,2)$
is the mass of the $i$-th constituent meson. $k$ is the momentum
of the exchanged mediator.

A detailed analysis on the Lorentz structure
\cite{Yu:2006ty,Guo:2007mm,Feng:2011zzb} determines the form of the B-S wave function of
the bound state comprising a vector and an axial vector mesons
( $D^*_s$ and $\bar D_{s1}$) in $S-$wave  as
\begin{eqnarray} \label{4-dim-BS1}
\langle0|T\phi_a(x_1)\phi_b(x_2)|V\rangle=\frac{\varepsilon_{abcd}}{\sqrt{6}M}\chi^d_P(x_1,x_2)P^c
,
\end{eqnarray}
where $a, b, c$ and $d$ are Lorentz indices. The wave function in
the momentum space can be obtained  by carrying out a Fourier transformation
\begin{eqnarray} \label{4-dim-BS3}
\chi^a_P(p_1,p_2)=\int d^4x_1 d^4x_2
e^{ip_1x_1+ip_2x_2}\chi^a_P(x_1,x_2)
=(2\pi)^4\delta(p_1+p_2+P)\chi^a_P(p).
\end{eqnarray}

Using the so-called ladder approximation one can get the
 B-S equation deduced in earlier references\cite{Salpeter:1952ib,Chang:2004im,Chang:2005sd}
\begin{eqnarray} \label{4-dim-BS4}
\varepsilon_{abcd}\chi^d_P(p)P^c
=\Delta_{1a\alpha}\int{d^4{q}\over(2\pi)^4}\,K^{\alpha\beta\mu\nu}(P,p,q)
\varepsilon_{\mu\nu\omega\sigma}\chi^\sigma_{P}(q)P^\omega\Delta_{2b\beta}\,,
\end{eqnarray}
where $\Delta_{1a\alpha}$ and $\Delta_{2b\beta}$ are the
propagators of $D^*_s$ and  $\bar D_{s1}$ respectively,
$K^{\alpha\beta\mu\nu}(P,p,q)$ is the kernel  determined by the
effective interaction between two constituents which can be
calculated from the Feynman diagrams in Fig. \ref{t1a} and
\ref{t1b}. In order to solve the B-S equation, we decompose the
relative momentum $p$ into the longitudinal component $p_l$ ($=
p\cdot v$) and the transverse one $p^\mu_t$ ($=
p^\mu-p_lv^\mu$)=(0, $\mathbf{p}_T$) with respect to  the momentum
of the bound state $P$ ($P=Mv$).

\begin{eqnarray}\label{propagator1}
\Delta^{a\alpha}_{1}=\frac{i[-g^{a\alpha}+p^a_1p^\alpha_1/
m^2_1]}{(\eta_1M+p_l+\omega_l-i\epsilon)(\eta_1M+p_l-\omega_l+i\epsilon)},
\end{eqnarray}
\begin{eqnarray}\label{propagator2}
\Delta^{b\beta}_{2}=\frac{i[-g^{b\beta}+p^b_2p^\beta_2/
m^2_2)]}{(\eta_2M-p_l+\omega_2-i\epsilon)(\eta_2M-p_l-\omega_2+i\epsilon)},
\end{eqnarray}
where $M$ is the mass of the bound state $Y(4626)$, $\omega_i =
\sqrt{{ \mathbf{p}_T}^2 + m_i^2}$.

By the Feynman diagrams shown in Fig. \ref{t1a} and \ref{t1b}, the
kernel $K^{\alpha\beta\mu\nu}(P,p,q)$ is written as
\begin{eqnarray}\label{kernel}
K^{\alpha\beta\mu\nu}(P,p,q)=&&\mathcal{C}_1g_{_{D_{s1}D^*_{s}\eta}}g_{_{\bar
D_{s1}\bar D^*_{s}\eta}}(\sqrt{6}k^\mu
k^\alpha-\sqrt{\frac{2}{3}}k^2g^{\mu\alpha}+\sqrt{\frac{2}{3}}k\cdot
p_1 k\cdot q_1 g^{\mu\alpha}/m_1/m_1')\nonumber\\&&
(\sqrt{6}k^\beta
k^\nu-\sqrt{\frac{2}{3}}k^2g^{\beta\nu}+\sqrt{\frac{2}{3}}k\cdot
p_2 k\cdot q_2
g^{\beta\nu}/m_2/m_2')\Delta(k,m_\eta)F^2(k,m_\eta)\nonumber\\&&-\frac{2}{3}\mathcal{C}_2g_{_{D^*_{s}D^*_{s}\eta}}g_{_{\bar
D_{s1}\bar D_{s1}\eta}}
\varepsilon^{\sigma\mu\alpha\omega}k_\sigma(p_{1\omega}+q_{1\omega})
\varepsilon^{\theta\nu\beta\rho}k_\theta(p_{2\rho}+q_{2\rho})\Delta(k,m_\eta)F^2(k,m_\eta)
\nonumber\\&&+\mathcal{C}_1g_{_{D_{s1}D^*_{s}\eta'}}g_{_{\bar
D_{s1}\bar D^*_{s}\eta'}}(\sqrt{6}k^\mu
k^\alpha-\sqrt{\frac{2}{3}}k^2g^{\mu\alpha}+\sqrt{\frac{2}{3}}k\cdot
p_1 k\cdot q_1 g^{\mu\alpha}/m_1/m_1')\nonumber\\&&
(\sqrt{6}k^\beta
k^\nu-\sqrt{\frac{2}{3}}k^2g^{\beta\nu}+\sqrt{\frac{2}{3}}k\cdot
p_2 k\cdot q_2
g^{\beta\nu}/m_2/m_2')\Delta(k,m_\eta')F^2(k,m_\eta')\nonumber\\&&-\frac{2}{3}\mathcal{C}_2g_{_{D^*_{s}D^*_{s}\eta'}}g_{_{\bar
D_{s1}\bar D_{s1}\eta'}}
\varepsilon^{\sigma\mu\alpha\omega}k_\sigma(p_{1\omega}+q_{1\omega})
\varepsilon^{\theta\nu\beta\rho}k_\theta(p_{2\rho}+q_{2\rho})\Delta(k,m_\eta')F^2(k,m_\eta')
\nonumber\\&&+\mathcal{C}_2[g_{_{D^*_{s}D^*_{s}\phi}}(q_1+p_1)^\chi
g^{\alpha\mu}-2g'_{_{D^*_{s}D^*_{s}\phi}}(k^\alpha
g^{\chi\mu}-k^\mu g^{\chi\alpha})](-g_{\chi \gamma }+k_\chi
k_\gamma /m^2_{\phi})\Delta(k,m_{\phi})\nonumber\\&&[g_{_{\bar
D_{s1}\bar D_{s1}\phi}}(q_1+p_1)^\gamma g^{\beta\nu}-2g'_{_{\bar
D_{s1}\bar D_{s1}\phi}}(k^\alpha g^{\gamma\mu}-k^\mu
g^{\gamma\alpha})]\nonumber\\&&+\mathcal{C}_1g_{_{D_{s1}
D^*_{s}\phi}}g_{_{\bar D_{s1} \bar
D^*_{s}\phi}}\varepsilon^{\alpha\mu\omega
\chi}(p_1+q_1)_\omega\varepsilon^{\beta\nu\rho \gamma
}(p_2+q_2)_\rho(-g_{\chi \gamma }+k_\chi k_\gamma
/m^2_{\phi})\Delta(k,m_{\phi})\nonumber\\&&+
\frac{2}{3}\mathcal{C}_2g_{_{D^*_{s}D^*_{s}f_0}}g_{_{\bar
D_{s1}\bar D_{s1}f_0}}g^{\mu\alpha}g^{\beta\nu}
\Delta(k,m_{f_0})F^2(k,m_{f_0}),
\end{eqnarray}
where $m_{\eta(\eta',\phi,f_0)}$ is the mass of the exchanged
meson $\eta(\eta',\phi(1020),f_0(980))$, $\mathcal{C}_1$=1 for
$Y_1$ and -1 for $Y_2$, $\mathcal{C}_2$=1,
$g_{_{D_{s1}D^*_{s}\eta}}$, $g_{_{\bar D_{s1}\bar D^*_{s}\eta}}$,
$g_{_{D^*_{s}D^*_{s}\eta}}$, $g_{_{\bar D_{s1}\bar
D_{s1}\eta}}$,$g_{_{D_{s1}D^*_{s}\eta'}}$, $g_{_{\bar D_{s1}\bar
D^*_{s}\eta'}}$, $g_{_{D^*_{s}D^*_{s}\eta'}}$, $g_{_{\bar
D_{s1}\bar D_{s1}\eta'}}$, $g_{_{D_{s1}D^*_{s}\phi}}$, $g_{_{\bar
D_{s1}\bar D^*_{s}\phi}}$, $g_{_{D^*_{s}D^*_{s}\phi}}$, $g_{_{\bar
D_{s1}\bar D_{s1}\phi}}$, $g'_{_{D^*_{s}D^*_{s}\phi}}$,
$g'_{_{\bar D_{s1}\bar D_{s1}\phi}}$, $g_{_{D^*_{s}D^*_{s}f_0}}$
and $g_{_{\bar D_{s1}\bar D_{s1}f_0}}$ are the concerned coupling
constants and $\Delta(k,m)=i/(k^2-m^2)$. Due to the small coupling
constants at the vertices  the contribution of $f_0(980)$ in
Fig.\ref{t1a} (b) is suppressed compared with that in
Fig.\ref{t1a} (a), so that we ignore the contribution of
$f_0(980)$  in Eq. (\ref{kernel}). All effective interactions are collected in Appendix.

Since the two constituents of the molecular state are not on
shell, at each interaction vertex a form factor should be
introduced to compensate the off-shell effect. The form factor is
employed in many references
\cite{Meng:2007tk,Cheng:2004ru,Liu:2006df,Ke:2010aw}, even
though it has different forms. Here we set it as:
\begin{eqnarray} \label{form-factor} F(k,m) = {\Lambda^2 -
m^2 \over \Lambda^2 + {\bf k}^2},
\end{eqnarray}
where ${\bf k}$ is the three-momentum of the exchanged meson and
$\Lambda$ is a cutoff parameter. Indeed, the form factor is introduced
phenomenologically and there lacks any reliable knowledge on the value of the cutoff parameter
$\Lambda$. $\Lambda$ is often parameterized to be
$\lambda\Lambda_{QCD}+m_s$ with $\Lambda_{QCD}=220$ MeV which is
adopted in some references
\cite{Meng:2007tk,Cheng:2004ru,Liu:2006df,Ke:2010aw}. As suggested, the
order of magnitude of the dimensionless parameter $\lambda$ should be close to 1.
In our later numerical computations, we set it to be within a wider range of $0\sim 4$ .

The wave function can be written as
\begin{eqnarray} \label{VS-BS}
\chi^d_{{P}}({ p}) =f({ p})\epsilon^d,
\end{eqnarray}
where $\epsilon$ is the polarization vector of the bound state and
$f({ p})$ is the radial wave function. The three-dimension spatial
wave function is obtained after integrating over $p_l$
\begin{eqnarray} \label{3-dim-BS1}
f({ |\mathbf{p_T}|}) =\int\frac{dp_l}{2\pi}f({ p}).
\end{eqnarray}

Substituting Eqs. (\ref{kernel}) and (\ref{VS-BS}) into Eq.
(\ref{4-dim-BS4}) and multiplying
$\varepsilon_{abfg}\chi^{*g}_P(x_1,x_2)P^f$ on both sides one can
sum over the polarizations of both sides.  Employing the so-called
covariant instantaneous approximation\cite{Dai:1993qr} $q_l=p_l$ i.e. using $p_l$ to
replace $q_l$ in $K(P,p,q)$, the kernel $K(P,p,q)$ does not
depend on $q_1$ any longer. Then we take a typical procedure:
integrating over $q_l$ on the
right side of Eq. (\ref{4-dim-BS4}), multiplying
$\int\frac{dp_l}{(2\pi)}$ on the both sides of Eq.
(\ref{4-dim-BS4}), and integrating over $p_l$ on the left side,
to reduce the expression into a compact form.
Finally we obtain
\begin{eqnarray} \label{four-dimension4}
&&6M^2f(|\mathbf{p}_T|)=\int\frac{dp_l}
{(2\pi)}\int\frac{d^3\mathbf{q}_T}{(2\pi)^3}\frac{
f(|\mathbf{q}_T|)}
{[(\eta_1M+p_l)^2-\omega^2_1+i\epsilon][(\eta_2M-p_l)^2-\omega^2_2+i\epsilon)]}\nonumber\\&&
[\mathcal{C}_1g_{_{D_{s1}D^*_{s}\eta}}g_{_{\bar D_{s1}\bar
D^*_{s}\eta}}F^2(k,m_\eta)\frac{C_0+C_1\,\mathbf{p}_T\cdot\mathbf{q}_T
+C_2(\mathbf{p}_T\cdot\mathbf{q}_T)^2+C_3(\mathbf{p}_T\cdot\mathbf{q}_T)^3+C_4(\mathbf{p}_T\cdot\mathbf{q}_T)^4}{-(\mathbf{p}_T-\mathbf{q}_T)^2-m_\eta^2}
\nonumber\\&&-\mathcal{C}_2g_{_{D^*_{s}D^*_{s}\eta}}g_{_{\bar
D_{s1}\bar
D_{s1}\eta}}F^2(k,m_\eta)\frac{C'_0+C'_1\,\mathbf{p}_T\cdot\mathbf{q}_T
}{-(\mathbf{p}_T-\mathbf{q}_T)^2-m_{\eta}^2}+\frac{\mathcal{C}_2g_{_{D^*_{s}D^*_{s}f_0}}g_{_{\bar
D_{s1}\bar D_{s1}f_0}}C_{S0}
}{-(\mathbf{p}_T-\mathbf{q}_T)^2-m_{f_0}^2}F^2(k,m_{f_0})\nonumber\\&&+\mathcal{C}_1g_{_{D_{s1}D^*_{s}\eta'}}g_{_{\bar
D_{s1}\bar D^*_{s}\eta'}}F^2(k,m_\eta')
\frac{C_0+C_1\,\mathbf{p}_T\cdot\mathbf{q}_T
+C_2(\mathbf{p}_T\cdot\mathbf{q}_T)^2+C_3(\mathbf{p}_T\cdot\mathbf{q}_T)^3+C_4(\mathbf{p}_T\cdot\mathbf{q}_T)^4}{-(\mathbf{p}_T-\mathbf{q}_T)^2-m_{\eta'}^2}
\nonumber\\&&-\mathcal{C}_2g_{_{D^*_{s}D^*_{s}\eta'}}g_{_{\bar
D_{s1}\bar
D_{s1}\eta'}}F^2(k,m_{\eta'})\frac{C'_0+C'_1\,\mathbf{p}_T\cdot\mathbf{q}_T
}{-(\mathbf{p}_T-\mathbf{q}_T)^2-m_{\eta'}^2}+\mathcal{C}_2F^2(k,m_{\phi})\frac{C'_{V0}+C'_{V1}\,\mathbf{p}_T\cdot\mathbf{q}_T+C'_{V2}\,(\mathbf{p}_T\cdot\mathbf{q}_T)^2
}{-(\mathbf{p}_T-\mathbf{q}_T)^2-m_{\phi}^2}\nonumber\\&&+\mathcal{C}_1g_{_{D_{s1}D^*_{s}\phi}}g_{_{\bar
D_{s1}\bar D^*_{s}\phi}}F^2(k,m_\phi)
\frac{C_{V0}+C_{V1}\,\mathbf{p}_T\cdot\mathbf{q}_T
+C_{V2}(\mathbf{p}_T\cdot\mathbf{q}_T)^2}{-(\mathbf{p}_T-\mathbf{q}_T)^2-m_{\phi}^2}],
\end{eqnarray}
with
\begin{eqnarray*}
C_0=&&4M^2{\left( {{\mathbf{p}_T}}^2 + {{\mathbf{q}_T}}^2 \right)
}^2+\frac{2M^2\left( {{m_1}}^2 + {{m_2}}^2 \right)
{{\mathbf{p}_T}}^2
    \left( 4{{\mathbf{p}_T}}^4 + 5{{\mathbf{p}_T}}^2{{\mathbf{q}_T}}^2 + {{\mathbf{q}_T}}^4 \right) }{3{{m_1}}^2{{m_2}}^2}
+\\&&\frac{2M^2{{\mathbf{p}_T}}^4{{\mathbf{q}_T}}^2\left(
-6{m_1}{m_2}{{\mathbf{q}_T}}^2 +
      {{m_1}}^2\left( -2{{\mathbf{p}_T}}^2 + {{\mathbf{q}_T}}^2 \right)  + {{m_2}}^2\left( -2{{\mathbf{p}_T}}^2 + {{\mathbf{q}_T}}^2 \right)  \right) }
    {3{{m_1}}^3{{m_2}}^3}\\&&-\frac{4M^2\left( {{m_1}}^2 + {{m_2}}^2 \right) {{\mathbf{p}_T}}^6{{\mathbf{q}_T}}^4}{3{{m_1}}^4{{m_2}}^4}
,
\\C_1=&&-16M^2\left( {{\mathbf{p}_T}}^2 + {{\mathbf{q}_T}}^2
\right)+\frac{-4M^2\left( {{m_1}}^2 + {{m_2}}^2 \right)
{{\mathbf{p}_T}}^2\left( 8{{\mathbf{p}_T}}^2 + 5{{\mathbf{q}_T}}^2
\right) }
  {3{{m_1}}^2{{m_2}}^2}+\\&&\frac{2M^2{{\mathbf{p}_T}}^2\left[ 12{m_1}{m_2}{{\mathbf{q}_T}}^2\left( {{\mathbf{p}_T}}^2 + {{\mathbf{q}_T}}^2 \right)  +
      ({{m_1}}^2+ {{m_2}}^2)\left( 2{{\mathbf{p}_T}}^4 + 5{{\mathbf{p}_T}}^2{{\mathbf{q}_T}}^2 - {{\mathbf{q}_T}}^4 \right)\right ]}{3{{m_1}}^3
    {{m_2}}^3}\\&&\frac{8M^2\left( {{m_1}}^2 + {{m_2}}^2 \right) {{\mathbf{p}_T}}^4{{\mathbf{q}_T}}^2\left( {{\mathbf{p}_T}}^2 + {{\mathbf{q}_T}}^2 \right) }
  {3{{m_1}}^4{{m_2}}^4}
,\\
  C_2=&&\frac{2M^2\left( {{m_2}}^2\left( 19{{\mathbf{p}_T}}^2 +
3{{\mathbf{q}_T}}^2 \right) +
 {{m_1}}^2\left( 24{{m_2}}^2 + 19{{\mathbf{p}_T}}^2 + 3{{\mathbf{q}_T}}^2 \right)  \right) }{3{{m_1}}^2{{m_2}}^2}
\\&&-\frac{4M^2\left[ ({{m_1}}^2+{{m_2}}^2){{\mathbf{p}_T}}^2\left(
{{\mathbf{p}_T}}^2 + {{\mathbf{q}_T}}^2 \right)    +
      {m_1}{m_2}\left( {{\mathbf{p}_T}}^4 + 4{{\mathbf{p}_T}}^2{{\mathbf{q}_T}}^2 + {{\mathbf{q}_T}}^4 \right)  \right] }{{{m_1}}^3
    {{m_2}}^3}\\&&-\frac{4M^2\left( {{m_1}}^2 + {{m_2}}^2 \right) {{\mathbf{p}_T}}^2
    \left( {{\mathbf{p}_T}}^4 + 4{{\mathbf{p}_T}}^2{{\mathbf{q}_T}}^2 + {{\mathbf{q}_T}}^4 \right) }{3{{m_1}}^4{{m_2}}^4},\\
  C_3=&&
 4M^2\left( -{{m_1}}^{-2} - {{m_2}}^{-2}
\right)+\frac{2M^2\left[ 12{m_1}{m_2}\left( {{\mathbf{p}_T}}^2 +
{{\mathbf{q}_T}}^2 \right)  +
      ({{m_1}}^2+{{m_2}}^2)\left( 7{{\mathbf{p}_T}}^2 + 3{{\mathbf{q}_T}}^2 \right)
      \right] }{3{{m_1}}^3{{m_2}}^3}\\&&+\frac{8M^2\left( {{m_1}}^2 + {{m_2}}^2 \right) {{\mathbf{p}_T}}^2\left( {{\mathbf{p}_T}}^2 + {{\mathbf{q}_T}}^2 \right) }
  {3{{m_1}}^4{{m_2}}^4}
,\\
  C_4=&&\frac{-2M^2\left[ 3{{m_1}}^3{m_2} + 3{m_1}{{m_2}}^3 +
2{{m_2}}^2{{\mathbf{p}_T}}^2 +
      2{{m_1}}^2\left( 3{{m_2}}^2 + {{\mathbf{p}_T}}^2 \right)  \right] }{3{{m_1}}^4{{m_2}}^4}\\C'_0=&&\frac{-16M^2\left( \eta_2M -p_l \right) \left( \eta_1M +p_l \right)
    \left( \mathbf{p}_T^2 + \mathbf{q}_T^2 \right) }{3m_1m_2},\\
    C'_1=&&\frac{32M^2\left( \eta_2M -p_l \right) \left( \eta_1M +p_l \right)
    }{3m_1m_2}\\C_{S0}=&&
\frac{-2\,M^2\,\left( {{m_1}}^2 + {{m_2}}^2 \right)
\,{{pt}}^2\,}{{{m_1}}^2\,{{m_2}}^2}-6\,M^2\\C_{V0}=&&-2M^2\left(
12{\eta_1}M\left( {\eta_2}M - {p_l} \right) + 12{\eta_2}M{p_l} -
12{{p_l}}^2 +
    {{\mathbf{p}_T}}^2 + {{\mathbf{q}_T}}^2 \right)\\&&-\frac{8M^2\left( {\eta_2}M - {p_l} \right) \left( {\eta_1}M + {p_l} \right)
    \left( {{\mathbf{p}_T}}^2 + {{\mathbf{q}_T}}^2 \right)
    }{{{m_v}}^2}\\&&- \frac{M^2{{\mathbf{p}_T}}^2\left[ 4{\eta_1}M\left( {\eta_2}M - {p_l} \right)  + 4{\eta_2}M{p_l} -
        4{{p_l}}^2 + {{\mathbf{q}_T}}^2 \right] }{{{m_1}}^2}
        \\&&- \frac{M^2{{\mathbf{p}_T}}^2\left[ 4{\eta_1}M\left( {\eta_2}M - {p_l} \right)  + 4{\eta_2}M{p_l} -
        4{{p_l}}^2 + {{\mathbf{q}_T}}^2 \right] }{{{m_2}}^2}
        ,
\\C_{V1}=&&-4M^2+\frac{16M^2\left(
{\eta_2}M - {p_l} \right) \left( {\eta_1}M + {p_l} \right)
}{{{m_v}}^2}+\frac{4M^2\left( {\eta_2}M - {p_l} \right) \left(
{\eta_1}M + {p_l} \right) }{{{m_1}}^2}\\&&+\frac{4M^2\left(
{\eta_2}M - {p_l} \right)
\left( {\eta_1}M + {p_l} \right) }{{{m_2}}^2},\\
C_{V2}=&&M^2\left( {{m_1}}^{-2} + {{m_2}}^{-2} \right),
\\C_{V0}'=&&\frac{8(g'_{_{D^*D^*\phi}}g_{_{\bar D_{s1}\bar
D_{s1}\phi}}m_2^2-g_{_{D^*D^*\phi}}g'_{_{\bar D_{s1}\bar
D_{s1}\phi}}m_1^2)M^2\left( {\eta_2}M - {p_l} \right) \left(
{\eta_1}M + {p_l} \right) {{\mathbf{p}_T}}^2}
  {{{m_1}}^2m_2^2}\\&&+\frac{-4g'_{_{D^*D^*\phi}}g'_{_{\bar D_{s1}\bar D_{s1}\phi}}M^2\left[({{m_1}}^2+ {{m_2}}^2){{\mathbf{p}_T}}^2\left( 2{{\mathbf{p}_T}}^2 + {{\mathbf{q}_T}}^2 \right)  +
      4 {{m_1}}^2{{m_2}}^2\left( {{\mathbf{p}_T}}^2 + {{\mathbf{q}_T}}^2 \right)
            \right]
         }{{{m_1}}^2{{m_2}}^2}\\&&+6g_{_{D^*D^*\phi}}g_{_{\bar D_{s1}\bar D_{s1}\phi}}M^2\left[ 4{\eta_1}M\left( {\eta_2}M - {p_l} \right)  + 4{\eta_2}M{p_l} -
    4{{p_l}}^2 + {{\mathbf{p}_T}}^2 + {{\mathbf{q}_T}}^2
    \right]\\&&+
\frac{6g_{_{D^*D^*\phi}}g_{_{\bar D_{s1}\bar
D_{s1}\phi}}M^2{\left( {{\mathbf{p}_T}}^2 - {{\mathbf{q}_T}}^2
\right) }^2}{{{m_v}}^2}
\\&&
+\frac{2g_{_{D^*D^*\phi}}g_{_{\bar D_{s1}\bar
D_{s1}\phi}}M^2{{\mathbf{p}_T}}^2\left[ 4{\eta_1}M\left( {\eta_2}M
- {p_l} \right)  +
      4{\eta_2}M{p_l} - 4{{p_l}}^2 + {{\mathbf{p}_T}}^2 + {{\mathbf{q}_T}}^2 \right]({{m_1}}^2+{{m_2}}^2) }{{{m_1}}^2{{m_2}}^2}
\\&&+\frac{2g_{_{D^*D^*\phi}}g_{_{\bar D_{s1}\bar D_{s1}\phi}}M^2{{\mathbf{p}_T}}^2{\left( {{\mathbf{p}_T}}^2 - {{\mathbf{q}_T}}^2 \right) }^2({{m_1}}^2+{{m_2}}^2)}{{{m_1}}^2{{m_2}}^2{{m_v}}^2}
\\C_{V1}'=&&\frac{8(g_{_{D^*D^*\phi}}g'_{_{\bar D_{s1}\bar D_{s1}\phi}}m_1^2-g'_{_{D^*D^*\phi}}g_{_{\bar D_{s1}\bar
D_{s1}\phi}}m_2^2)M^2\left( {\eta_2}M - {p_l} \right) \left(
{\eta_1}M + {p_l} \right)
}{m_1^2{{m_2}}^2}
\\&&+4g_{_{D^*D^*\phi}}g_{_{\bar D_{s1}\bar D_{s1}\phi}}M^2\left( 3 + \frac{{{\mathbf{p}_T}}^2}{{{m_1}}^2} + \frac{{{\mathbf{p}_T}}^2}{{{m_2}}^2}
\right)+16g'_{_{D^*D^*\phi}}g'_{_{\bar D_{s1}\bar
D_{s1}\phi}}M^2\left( 2 + \frac{{{\mathbf{p}_T}}^2}{{{m_1}}^2} +
\frac{{{\mathbf{p}_T}}^2}{{{m_2}}^2} \right)
\\C_{V2}'=&&\frac{-4g'_{_{D^*D^*\phi}}g'_{_{\bar D_{s1}\bar D_{s1}\phi}}M^2\left( {{m_1}}^2 + {{m_2}}^2 \right)
}{{{m_1}}^2{{m_2}}^2}
\end{eqnarray*}

While we integrate over $p_l$ on the right side of Eq.
(\ref{four-dimension4}) there exist four poles which are located at
$-\eta_1M-\omega_1+i\epsilon$, $-\eta_1M+\omega_1-i\epsilon$,
$\eta_2M+\omega_2-i\epsilon$ and $\eta_2M-\omega_2+i\epsilon$. By
choosing an appropriate contour we only need to evaluate the
residuals at $p_l=-\eta_1M-\omega_1+i\epsilon$ and
$p_l=\eta_2M-\omega_2+i\epsilon$.
%\begin{eqnarray}\label{residual}
%\int^{\infty}_{-\infty}\frac{dp_l} {(2\pi)}\frac{1 }
%{[(\eta_1M+p_l)^2-\omega^2_l+i\epsilon][(\eta_1M-p_l)^2-\omega^2_2+i\epsilon)]}=\frac{-i(\omega_1+\omega_2)}{2\omega_1\omega_2[M^2-(\omega_1+\omega_2)^2]}.
%\end{eqnarray}

Here $d^3\mathbf{q}_T=\mathbf{q}_T^2{\rm
sin}(\theta)d|\mathbf{q}_T|d\theta d\phi$ and $\mathbf{p}_T\cdot
\mathbf{q}_T=|\mathbf{p}_T||\mathbf{q}_T|{\rm cos}(\theta)$,
one can integrate out the
azimuthal part  and then Eq.
(\ref{four-dimension4}) is reduced into a one-dimensional integral
equation
\begin{eqnarray}\label{one dimension equation}
&&f(|\mathbf{p}_T|)=\int{\frac{|\mathbf{q}_T|^2f(|\mathbf{q}_T|)}{12M^2(2\pi)^2}d|\mathbf{q}_T|}\{
\frac{\mathcal{C}_1g_{_{D_{s1}D^*_{s}\eta}}g_{_{\bar D_{s1}\bar
D^*_{s}\eta}}(\omega_1+\omega_2)}{\omega_1\omega_2[M^2-(\omega_1+\omega_2)^2]}[C_0J_0(m_\eta)+C_1\,J_1(m_\eta)\nonumber\\&&
+C_2J_2(m_\eta)+C_3J_3(m_\eta)+C_4J_4(m_\eta)]-\frac{\mathcal{C}_2g_{_{D^*_{s}D^*_{s}\eta}}g_{_{\bar
D_{s1}\bar D_{s1}\eta}}} {\omega_1[(M+\omega_1)^2-\omega_2^2]}
[C'_0J_0(m_\eta)+C'_1\,J_1(m_\eta)]|_{p_l=-\eta_1M-\omega_1}\nonumber\\&&-\frac{\mathcal{C}_2g_{_{D^*_{s}D^*_{s}\eta}}g_{_{\bar
D_{s1}\bar D_{s1}\eta}}} {\omega_2[(M-\omega_2)^2-\omega_1^2]}
[C'_0J_0(m_\eta)+C'_1\,J_1(m_\eta)]|_{p_l=\eta_2M-\omega_2}
+\frac{\mathcal{C}_2g_{_{D^*_{s}D^*_{s}f_0}}g_{_{\bar D_{s1}\bar
D_{s1}f_0}}(\omega_1+\omega_2)}{\omega_1\omega_2[M^2-(\omega_1+\omega_2)^2]}C_{S0}J_0(m_{f_0})\nonumber\\&&+
\frac{\mathcal{C}_1g_{_{D_{s1}D^*_{s}\eta'}}g_{_{\bar D_{s1}\bar
D^*_{s}\eta'}}(\omega_1+\omega_2)}{\omega_1\omega_2[M^2-(\omega_1+\omega_2)^2]}[C_0J_0(m_\eta')+C_1\,J_1(m_\eta')
+C_2J_2(m_\eta')+C_3J_3(m_\eta')+C_4J_4(m_\eta')]
\nonumber\\&&-\frac{\mathcal{C}_2g_{_{D^*_{s}D^*_{s}\eta'}}g_{_{\bar
D_{s1}\bar D_{s1}\eta'}}} {\omega_1[(M+\omega_1)^2-\omega_2^2]}
[C'_0J_0(m_\eta)+C'_1\,J_1(m_\eta')]|_{p_l=-\eta_1M-\omega_1}\nonumber\\&&-\frac{\mathcal{C}_2g_{_{D^*_{s}D^*_{s}\eta'}}g_{_{\bar
D_{s1}\bar D_{s1}\eta'}}} {\omega_2[(M-\omega_2)^2-\omega_1^2]}
[C'_0J_0(m_\eta')+C'_1\,J_1(m_\eta')]|_{p_l=\eta_2M-\omega_2}\nonumber\\&&+\frac{\mathcal{C}_1g_{_{D_{s1}D^*_{s}\phi}}g_{_{\bar
D_{s1}\bar D^*_{s}\phi}}} {\omega_1[(M+\omega_1)^2-\omega_2^2]}
[C_{V0}J_0(m_\phi)+C_{V1}\,J_1(m_\phi)+C_{V2}\,J_2(m_\phi)]|_{p_l=-\eta_1M-\omega_1}
\nonumber\\&&+\frac{\mathcal{C}_1g_{_{D^*_{s}D^*_{s}\phi}}g_{_{\bar
D_{s1}\bar D_{s1}\phi}}} {\omega_2[(M-\omega_2)^2-\omega_1^2]}
[C_{V0}J_0(m_\phi)+C_{V1}\,J_1(m_\phi)+C_{V2}\,J_2(m_\phi)]|_{p_l=\eta_2M-\omega_2}\nonumber\\&&+\frac{\mathcal{C}_2}
{\omega_1[(M+\omega_1)^2-\omega_2^2]}
[C'_{V0}J_0(m_\phi)+C'_{V1}\,J_1(m_\phi)+C'_{V2}\,J_2(m_\phi)]|_{p_l=-\eta_1M-\omega_1}
\nonumber\\&&+\frac{\mathcal{C}_2}
{\omega_2[(M-\omega_2)^2-\omega_1^2]}
[C'_{V0}J_0(m_\phi)+C'_{V1}\,J_1(m_\phi)+C'_{V2}\,J_2(m_\phi)]|_{p_l=\eta_2M-\omega_2}
\},
\end{eqnarray}
with
\begin{eqnarray*}\label{J01234}
J_0(m)&&=\int^\pi_0\frac{{\rm
sin}\theta\,d\theta}{-(\mathbf{p}_T-\mathbf{q}_T)^2-m^2}F^2(k,m),
J_1(m)=\int^\pi_0\frac{|\mathbf{p}_T||\mathbf{q}_T|{\rm sin}\theta{\rm cos}\theta\,d\theta}{-(\mathbf{p}_T-\mathbf{q}_T)^2-m^2}F^2(k,m),\\
J_2(m)&&=\int^\pi_0\frac{|\mathbf{p}_T|^2|\mathbf{q}_T|^2{\rm
sin}\theta{\rm
cos^2}\theta\,d\theta}{-(\mathbf{p}_T-\mathbf{q}_T)^2-m^2}F^2(k,m),
J_3(m)=\int^\pi_0\frac{|\mathbf{p}_T|^3|\mathbf{q}_T|^3{\rm sin}\theta{\rm cos^3}\theta\,d\theta}{-(\mathbf{p}_T-\mathbf{q}_T)^2-m^2}F^2(k,m),\\
J_4(m)&&=\int^\pi_0\frac{|\mathbf{p}_T|^4|\mathbf{q}_T|^4{\rm
sin}\theta{\rm
cos^4}\theta\,d\theta}{-(\mathbf{p}_T-\mathbf{q}_T)^2-m^2}F^2(k,m).
\end{eqnarray*}

\subsection{Normalization condition for the B-S wave function}
In analog to the cases in Refs.\cite{Guo:2007mm,Feng:2011zzb} the
normalization condition for the B-S wave function of a bound state
should be
\begin{eqnarray}\label{normal1}
\frac{i}{6}\int
\frac{d^4pd^4q}{(2\pi)^8}\varepsilon_{abcd}\bar\chi^d_P(p)\frac{P^c}{M}\frac{\partial}{\partial
P_0}[I^{ab\alpha\beta}(P,p,q)+K^{ab\alpha\beta}(P,p,q)]\varepsilon_{\alpha\beta\mu\nu}\chi^\nu_P(q)\frac{P^\mu}{M}=1,
\end{eqnarray}
where $P_0$ is the energy of the bound state which is equal to its
mass $M$ in the center of mass frame. $I(P,p,q)$ is a product of
reciprocals of two free propagators with a proper weight.

\begin{eqnarray}
I^{ab\alpha\beta}(P,p,q)=(2\pi)^4\delta^4(p-q)(\Delta^{a\alpha}_1)^{-1}(\Delta^{b\beta}_2)^{-1}.
\end{eqnarray}
In our earlier work \cite{Ke:2012gm} we found that the term
$K^{ab\alpha\beta}(P,p,q)$ in brackets is negligible, so that now we ignore
it as done in Ref.\cite{Feng:2012zzf}.

To reduces the singularity of the problem  we ignore the second
item in the numerators of the propagators (Eq. (\ref{propagator1})
and (\ref{propagator2}) ) and
$(\Delta^{a\alpha}_1)^{-1}=-ig^{a\alpha}(p_1^2-m_1^2)$,
$(\Delta^{b\beta}_1)^{-1}=-ig^{b\beta}(p_2^2-m_2^2)$. The
normalization condition is
\begin{eqnarray}\label{normal2}
i\int \frac{d^4pd^4q}{(2\pi)^8}f^*(p)\frac{\partial}{\partial
P_0}[(2\pi)^4\delta^4(p-q)(p_1^2+m_1^2)(p_2^2+m_2^2)]f(q)=2M.
\end{eqnarray}

After performing some manipulations we obtain the normalization
of the radial wave function
\begin{eqnarray}\label{normal3}
\frac{1}{2M}\int \frac{d^3
\mathbf{p_T}}{(2\pi)^3}f^2(|\mathbf{p_T}|)\frac{M\omega_1\omega_2}{\omega_1+\omega_2}=1.
\end{eqnarray}

\section{The strong decays of the molecular state $Y(4626)$}

Now we investigate the strong decays of $Y(4626)$ using the
effective interactions which only includes contributions induced
by exchanging $\eta$ and $\eta'$. We will further discuss this
issue latter.
\begin{figure*}
        \centering
        \subfigure[~]{
          \includegraphics[width=7cm]{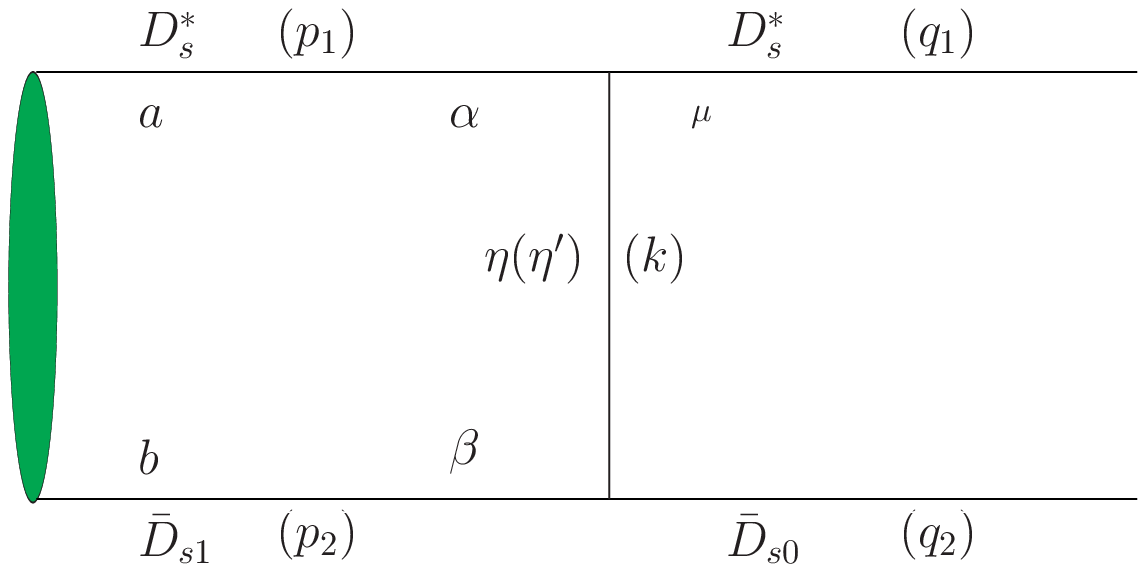}}\,\,\,\,\,\,\,\,\,\,\,\,\,\,\,\,\,\,\,\,\,\,
        \subfigure[~]{
          \includegraphics[width=7cm]{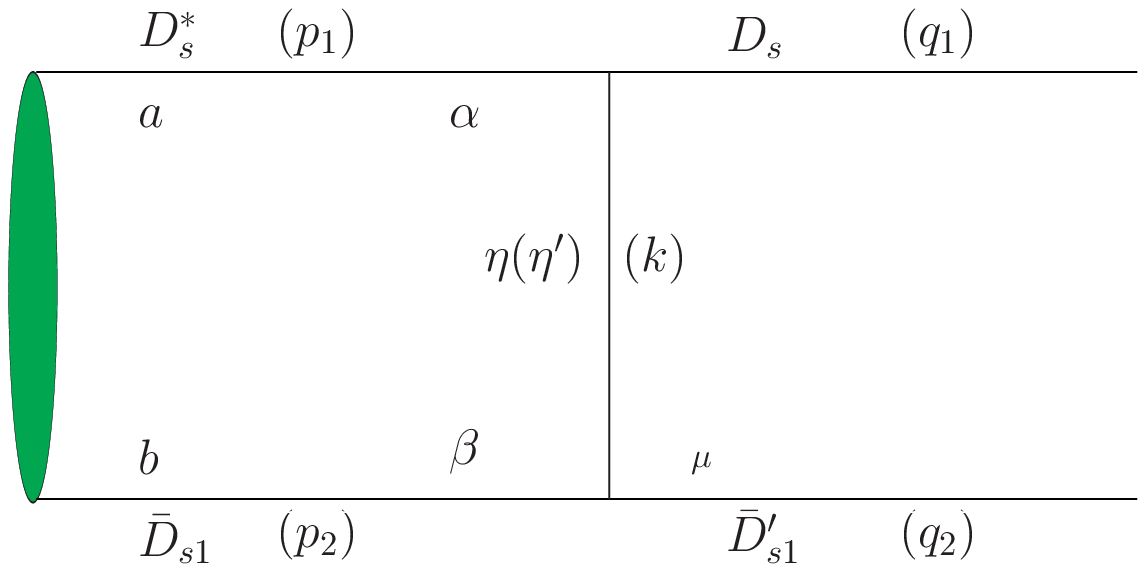}}\\  \subfigure[~]{
          \includegraphics[width=7cm]{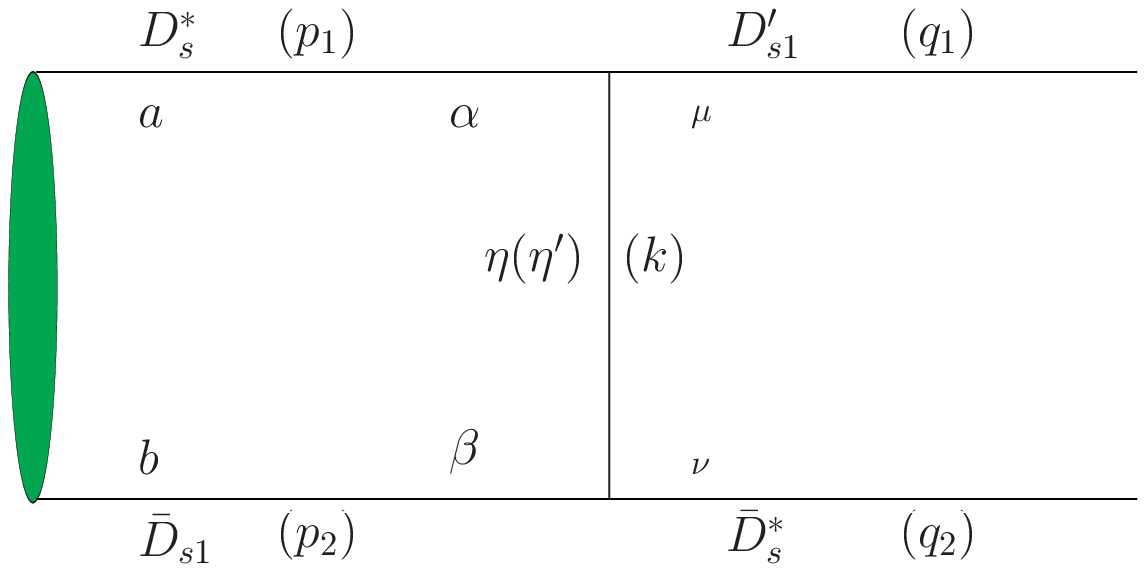}}\,\,\,\,\,\,\,\,\,\,\,\,\,\,\,\,\,\,\,\,\,\,
        \subfigure[~]{
          \includegraphics[width=7cm]{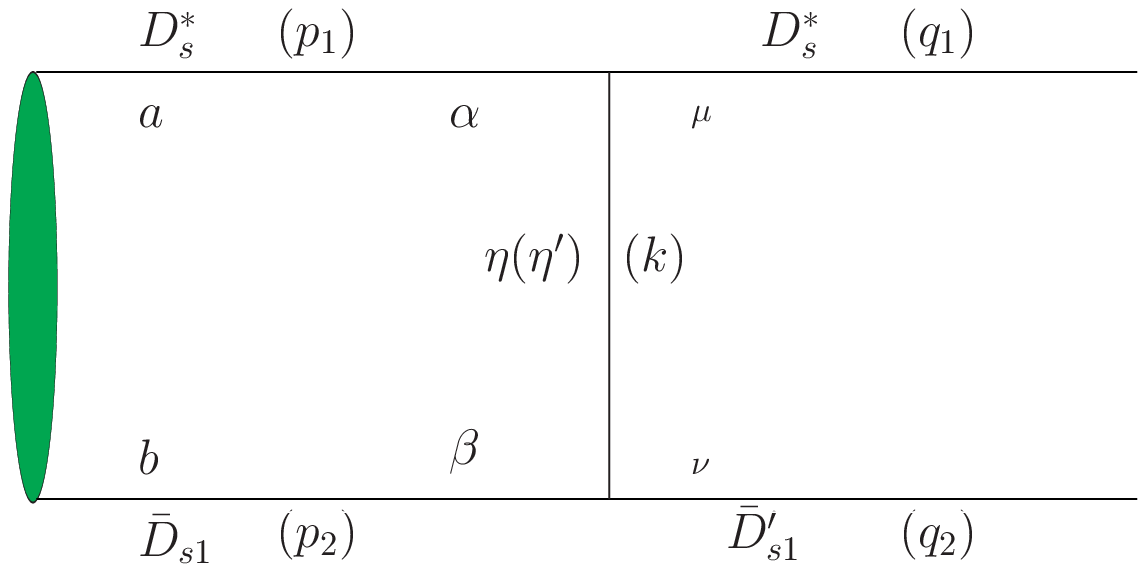}}\\  \subfigure[~]{
          \includegraphics[width=7cm]{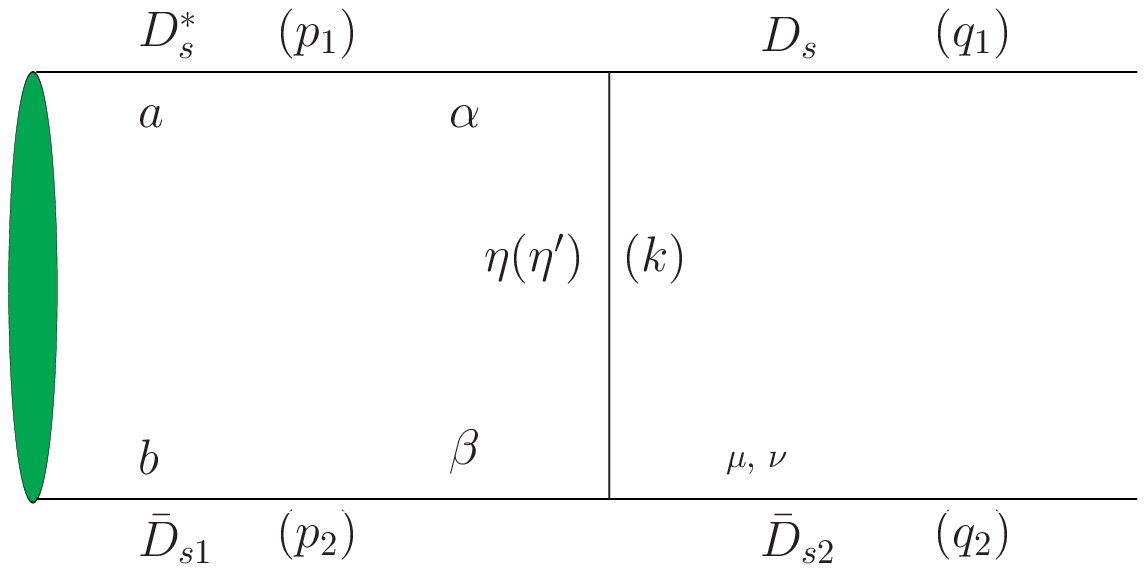}}\,\,\,\,\,\,\,\,\,\,\,\,\,\,\,\,\,\,\,\,\,\,
        \subfigure[~]{
          \includegraphics[width=7cm]{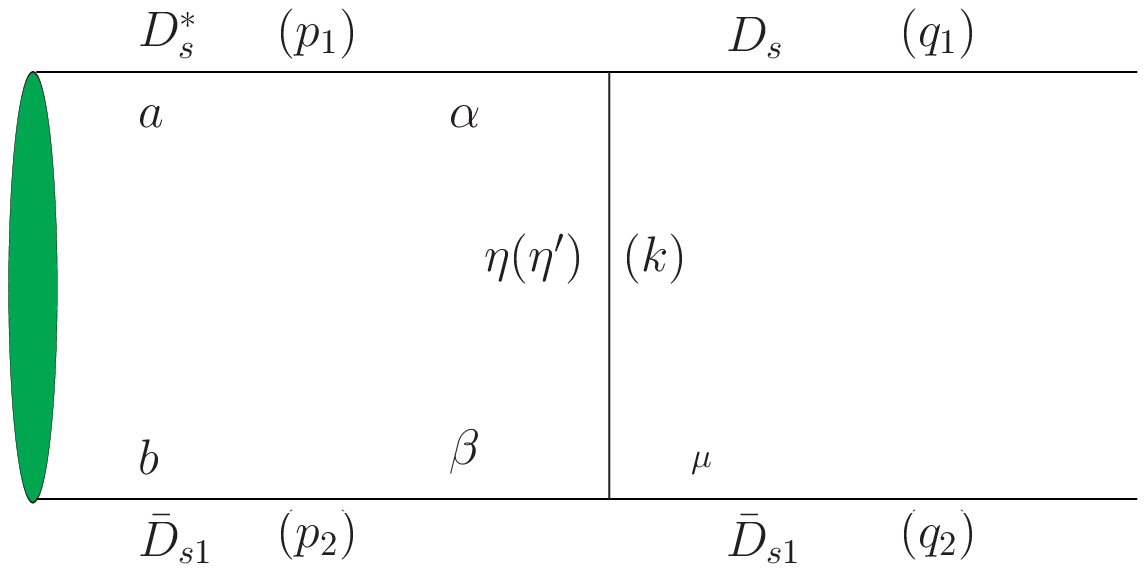}}
 \caption{the decays of  $Y(4626)$  by exchanging $\eta(\eta')$.}
        \label{decay}
    \end{figure*}

 \subsection{Decay to $ D_s^*(1^-)+\bar D_{s0}(2317)(0^+)$ }
The relevant Feynman diagram is depicted in Fig. \ref{decay} (a)
where $\bar D_{s0}$ represents $\bar D_{s0}(2317)$. The amplitude
is,
\begin{eqnarray}\label{a1}
\mathcal{A}_{a}=&&g_{_{D^*_{s}D^*_{s}\eta}}g_{_{\bar D_{s1}\bar
D_{s0}\eta}}\int\frac{d^4p}{(2\pi)^4}\frac{2}{3}k_{\nu}\epsilon_{1\mu}\varepsilon^{\nu\mu
a\beta}(\frac{p_{1\beta}}{m_1}+ \frac{q_{1\beta}}{m'_1})\bar
\chi^d(p)\varepsilon_{abcd}\frac{P^c}{M}k_b\Delta(k,m_\eta)F^2(k,m_\eta)\nonumber\\&&+{\rm
{a\,\,\,\, term\,\,\,\, with\,\,\,\, \eta'\,\,\,\, replacing
\,\,\,\,\eta }},
\end{eqnarray}
where $k=p-(\eta_2 q_1-\eta_1 q_2)$, $\epsilon_1$ is the
polarization vector of $D^*_s$. We still take the approximation
$k_0=0$ to carry out the calculation.

The amplitude can be parameterized as\cite{Chung:1993da}
\begin{eqnarray}\label{a3}
\mathcal{A}_a= g_0M\epsilon_1\cdot \epsilon^*+\frac{g_2}{M}(q\cdot
\epsilon_1 q\cdot \epsilon^*-\frac{1}{3}q^2\epsilon_1\cdot
\epsilon^*).
\end{eqnarray}
 The
factors $g_0$ and $g_2$ are extracted from the expressions of
$\mathcal{A}_a$.

 Then the partial width is expressed as
\begin{eqnarray}
d\Gamma_a=\frac{1}{32\pi^2}|\mathcal{A}_a|^2\frac{|q_2|}{M^2}d\Omega.
\end{eqnarray}

\subsection{Decay to $D_s(0^-)+\bar D_s(2460)(1^+)$ }

The corresponding Feynman diagram is depicted in Fig. \ref{decay} (b) where
$\bar D'_{s1}$ denotes $D_s(2460)$ through the whole paper. The
amplitudes is
\begin{eqnarray}\label{b1}
\mathcal{A}_{b}=&&g_{_{D^*_{s}D_{s}\eta}}g_{_{\bar D_{s1}\bar
D_{s(2460)}\eta}}\int\frac{d^4p}{(2\pi)^4}\frac{2}{3}k^{a}\bar
\chi^d(p)\varepsilon_{abcd}\frac{P^c}{M}\epsilon_{2\mu}\varepsilon^{\nu\mu
b\omega}(\frac{p_{1\omega}}{m_1}+
\frac{q_{1\omega}}{m'_1})k_\nu\Delta(k,m_\eta)F^2(k,m_\eta)\nonumber\\&&+{\rm
{a\,\,\,\, term\,\,\,\, with\,\,\,\, \eta'\,\,\,\, replacing
\,\,\,\,\eta}} .
\end{eqnarray}

The amplitude can also be parameterized as
\begin{eqnarray}\label{b3}
\mathcal{A}_b= g'_0M\epsilon_2\cdot
\epsilon^*+\frac{g'_2}{M}(q\cdot \epsilon_2 q\cdot
\epsilon^*-\frac{1}{3}q^2\epsilon_2\cdot \epsilon^*),
\end{eqnarray}
where $\epsilon_2$ is the polarizations of $\bar D_s(2460)$.
The factors $g'_0$ and $g'_2$ can be extracted from
the expressions of $\mathcal{A}_b$.

\subsection{Decay to $D_s(2460)(1^+)+\bar D^*_s(1^-)$ }

The Feynman diagram for the process of $Y(4626)\to
D_s(2460)(1^+)+\bar D^*_s(1^-)$ is depicted in Fig. \ref{decay}
(c). The amplitudes is
\begin{eqnarray}\label{c1}
\mathcal{A}_{c}=&&g_{_{D^*_{s}D_{s(2460)}\eta}}g_{_{\bar
D_{s1}\bar
D^*_{s}\eta}}\int\frac{d^4p}{(2\pi)^4}\frac{2}{3}ik_{\omega}(\frac{p^\omega_1}{m1}+
\frac{q^\omega_1}{m'_1})\epsilon^a_{1}\bar
\chi^d(p)\varepsilon_{abcd}\frac{P^c}{M}\nonumber\\&&(-3k^b
k^\nu+k^2g^{b\nu}-k\cdot p_2 k\cdot q_2
g^{b\nu}/m_2/m_2')\epsilon_{2\nu}\Delta(k,m_\eta)F^2(k,m_\eta)\nonumber\\&&+{\rm
{a\,\,\,\, term\,\,\,\, with\,\,\,\, \eta'\,\,\,\, replacing
\,\,\,\,\eta}},
\end{eqnarray}
where $\epsilon_1$ and $\epsilon_2$ are the polarization vectors of
$D_s(2460)$ and $\bar D^*_s$ respectively.
The total amplitude can be parameterized as\cite{Chung:1993da}
\begin{eqnarray}\label{c3}
\mathcal{A}_c=&&
g_{10}\varepsilon^{\mu\nu\alpha\beta}P_\mu\epsilon_{1\nu}\epsilon_{2\alpha}\epsilon^*_\beta
+\frac{g_{11}}{M^2}\varepsilon^{\mu\nu\alpha\beta}P_\mu
q_\nu\epsilon_{1\alpha}\epsilon_{2\beta}
 q\cdot \epsilon^*\nonumber\\&&+\frac{g_{12}}{M^2}\varepsilon^{\mu\nu\alpha\beta}P_\mu
q_\nu\epsilon_{1\alpha}\epsilon^*_{\beta}
 q\cdot \epsilon_2.
\end{eqnarray}
The factors $g_{10}$, $g_{11}$ and $g_{12}$ are extracted from
the expressions of $\mathcal{A}_c$.

\subsection{Decay to $D^*_s(1^-)+\bar D_s(2460)(1^+)$ }

The Feynman diagram for $Y(4626)\to D^*_s(1^-)+\bar
D_s(2460)(1^+)$ is depicted in Fig. \ref{decay} (d). The amplitude
is
\begin{eqnarray}\label{d1}
\mathcal{A}_{d}=&&g_{_{D^*_{s}D^*_{s}\eta}}g_{_{\bar D_{s1}\bar
D_{s(2460)}\eta}}\int\frac{d^4p}{(2\pi)^4}\frac{2}{3}k_{\sigma}\epsilon_{1\mu}\varepsilon^{\sigma
a\mu\gamma}(\frac{p^\gamma_1}{m_1}+ \frac{q^\gamma_2}{m'_2})\bar
\chi^d(p)\varepsilon_{abcd}\frac{P^c}{M}\nonumber\\&&k_\omega\epsilon_{2\nu}\varepsilon^{\omega\nu
b \theta}(\frac{p_{2\theta}}{m_2}+
\frac{q_{1\theta}}{m'_1})\Delta(k,m_\eta)F^2(k,m_\eta)+{\rm
{a\,\,\,\, term\,\,\,\, with\,\,\,\, \eta'\,\,\,\, replacing
\,\,\,\,\eta},}
\end{eqnarray}
where $\epsilon_1$ and $\epsilon_2$ are the polarization vectors of
$D^*_s$ and $\bar D_s(2460)$ respectively.

The total amplitude for the strong decay of $Y(4626)\to D^*_s(1^-)+\bar D_s(2460)(1^+)$  can also be
expressed as
\begin{eqnarray}\label{d3}
\mathcal{A}_d=&&
g'_{10}\varepsilon^{\mu\nu\alpha\beta}P_\mu\epsilon_{1\nu}\epsilon_{2\alpha}\epsilon^*_\beta
+\frac{g'_{11}}{M^2}\varepsilon^{\mu\nu\alpha\beta}P_\mu
q_\nu\epsilon_{1\alpha}\epsilon_{2\beta}
 q\cdot \epsilon^*\nonumber\\&&+\frac{g'_{12}}{M^2}\varepsilon^{\mu\nu\alpha\beta}P_\mu
q_\nu\epsilon_{1\alpha}\epsilon^*_{\beta}
 q\cdot \epsilon_2.
\end{eqnarray}
The factors $g'_{10}$, $g'_{11}$ and $g'_{12}$ are extracted
from the expressions of $\mathcal{A}_d$.

\subsection{Decay to $D_s(0^-)+\bar D_s(2572)(2^+)$ }

The Feynman diagram is depicted in Fig. \ref{decay}(e) where $\bar
D_{s2}$ represents $\bar D_s(2572)$. The amplitudes is,
\begin{eqnarray}\label{e1}
\mathcal{A}_{e}=&&g_{_{D^*_{s}D_{s}\eta}}g_{_{D_{s1}D_{s2}\eta}}\int\frac{d^4p}{(2\pi)^4}\frac{2}{3}k^{a}\bar
\chi^d(p)\varepsilon_{abcd}\frac{P^c}{M}k_\mu\epsilon_{2}^{b\mu}\Delta(k,m_s)F^2(k,m_s)\nonumber\\&&+{\rm
{a\,\,\,\, term\,\,\,\, with\,\,\,\, \eta'\,\,\,\, replacing
\,\,\,\,\eta}},
\end{eqnarray}
where $\epsilon_2$ is the polarization tensor of $\bar D_s(2572)(2^+)$.

The total amplitude is written as
\begin{eqnarray}\label{e3}
\mathcal{A}_e=&&
\frac{g_{20}}{M^2}\varepsilon^{\mu\nu\alpha\beta}P_\mu\epsilon_{2\nu\sigma}q_{\alpha}\epsilon^*_\beta
q^\sigma.
\end{eqnarray}
The factors $g_{20}$ can be extracted from the expressions of
$\mathcal{A}_e$.

\subsection{Decay to $D_s(0^-)+\bar D_s(2536)(1^+)$ }

The Feynman diagram is depicted in Fig. \ref{decay} (f) where
$\bar D_{s1}$ represents $\bar D_s(2536)$. The amplitudes is
\begin{eqnarray}\label{f1}
\mathcal{A}_{f}=&&g_{_{D^*_{s}D_{s}\eta}}g_{_{\bar D_{s1}\bar
D_{s1}\eta}}\int\frac{d^4p}{(2\pi)^4}\frac{2}{3}k^{a}\bar
\chi^d(p)\varepsilon_{abcd}\frac{P^c}{M}\epsilon_{2\mu}\varepsilon^{\nu\mu
b\omega}(\frac{p_{2\omega}}{m_2}+
\frac{q_{2\omega}}{m'_2})k_\nu\Delta(k,m_\eta)F^2(k,m_\eta)\nonumber\\&&+{\rm
{a\,\,\,\, item\,\,\,\, with\,\,\,\, \eta'\,\,\,\, replacing
\,\,\,\,\eta}},
\end{eqnarray}
where $\epsilon_2$ is the polarization vector of $D_s(2536)$.

The amplitude is still written as
\begin{eqnarray}\label{f3}
\mathcal{A}_b= g''_0M\epsilon_2\cdot
\epsilon^*+\frac{g''_2}{M}(q\cdot \epsilon_2 q\cdot
\epsilon^*-\frac{1}{3}q^2\epsilon_2\cdot \epsilon^*).
\end{eqnarray}
The factors $g''_0$ and $g''_2$ are extracted from the
expressions of $\mathcal{A}_f$.

\section{numerical results}
\subsection{the numerical results }

Before we numerically solve the B-S equation all necessary
parameters should be priori determined as accurate as possible.
The masses $m_{D^*_s}$, $m_{D_{s0}}$, $m_{D_{s1}}$, $m_{D'_{s1}}$,
$m_{D_{s2}}$, $m_\eta$, $m_{\eta'}$, $m_{f_0(980)}$ and $m_\phi$
come from the databook\cite{PDG18}. The coupling constants in the
effective interactions $g_{_{D_{s1}D^*_{s}\eta}}$, $g_{_{\bar
D_{s1}\bar D^*_{s}\eta}}$, $g_{_{D^*_{s}D^*_{s}\eta}}$, $g_{_{\bar
D_{s1}\bar D_{s1}\eta}}$,$g_{_{D_{s1}D^*_{s}\eta'}}$, $g_{_{\bar
D_{s1}\bar D^*_{s}\eta'}}$, $g_{_{D^*_{s}D^*_{s}\eta'}}$,
$g_{_{\bar D_{s1}\bar D_{s1}\eta'}}$, $g_{_{D_{s1}D^*_{s}\phi}}$,
$g_{_{\bar D_{s1}\bar D^*_{s}\phi}}$, $g_{_{D^*_{s}D^*_{s}\phi}}$,
$g_{_{\bar D_{s1}\bar D_{s1}\phi}}$, $g'_{_{D^*_{s}D^*_{s}\phi}}$,
$g'_{_{\bar D_{s1}\bar D_{s1}\phi}}$, $g_{_{D^*_{s}D^*_{s}f_0}}$
and $g_{_{\bar D_{s1}\bar D_{s1}f_0}}$ are taken from the relevant
literatures and their values and related references are collected
in the Appendix.

With these input parameters the B-S equation Eq. (\ref{one
dimension equation}) can be solved numerically. Since it is an
integral equation, an efficient way for solving it is discretizing
it and then turns solving the integral equation to an algebraic
equation group. Concretely, we let the variables $\bf |p_T|$ and
$\bf |q_T|$ be discretized into $n$ values $Q_1$, $Q_2$,...$Q_n$
(when $n>100$ the solution is stable enough, and we set $n$=129 in
our calculation) and the equal gap between two adjacent values as
$\frac{Q_n-Q_1}{n-1}$. Here we set $Q_1$=0.001 GeV and $Q_n$=2 GeV
. The $n$ values of $f({\bf |p_T|})$ constitute a column matrix on
the left side of the equation and the $n$ elements $f({\bf
|q_T|})$ construct another column matrix on the right side of the
equation as shown below. In this case, the functions in the curl
bracket of Eq. (\ref{one dimension equation}) multiplying
${\frac{|\mathbf{q}_T|^2}{12M^2(2\pi)^2}}$ would be an effective
operator acting on $f({\bf |q_T|})$. It is specially noted that
because discretizing the equation, even ${\frac{|\mathbf{q}_T|^2}
{12M^2(2\pi)^2}}$ turns from continuous integration variable into
$n$ discrete values which are involved in the $n\times n$
coefficient matrix. Substituting the $n$ pre-set $Q_i$ values into
those functions, the operator turns into an $n\times n$ matrix
which associates the two column matrices. It is noted that $Q_1$,
$Q_2$,...$Q_n$ should take sequential values.
$$\left(\begin{array}{c}
      f(Q_1) \\... \\f(Q_{129})
      \end{array}\right)=A(\Delta E, \lambda)\left(\begin{array}{c}
     f(Q_1) \\... \\f(Q_{129}))
      \end{array}\right).$$
As is well known, if a homogeneous equation possesses non-trivial
solutions, the necessary and sufficient condition is det$|A(\Delta
E,\lambda)-I|=0$ ($I$ is the unit matrix) where $A(\Delta
E,\lambda)$ is just the aforementioned coefficient matrix. Thus
solving the integral equation just turns to a sort of eigenvalue
searching problem which is a familiar issue in quantum mechanics,
in particular, the eigenvalue is required to be unit in this
problem. Here $A(\Delta E,\lambda)$ is a function of the binding
energy $\Delta E=m_1+m_2-M$ and parameter $\lambda$. The following
procedure is a bit tricky. Inputting a supposed $\Delta E$, we
vary $\lambda$ to make det$|A(\Delta E,\lambda)-I|=0$ hold.  One
can note that the matrix equation $(A(\Delta
E,\lambda)_{ij})(f(j))=\beta (f(i))$ is exactly an eigenequation.
Using the values of $\Delta E$ and  $\lambda$, we seek out all
possible ``eigenvalues" $\beta$. Among them only $\beta=1$ is the
solution we expect. In the process of solving the equation group,
the value of $\lambda$ is determined, and actually it is the
solution of the equation group with $\beta=1$. Meanwhile using the
obtained $\lambda$,  one achieve the corresponding wavefunction
$f(Q_1),f(Q_2)...f(Q_{129})$ which just is the solution of the B-S
equation.

Generally $\lambda$ should be within the range around the order of
unit. In Ref.\cite{Meng:2007tk} the authors fixed $\lambda$ to be
3. In our earlier paper\cite{Ke:2010aw}  the value of $\lambda$
varied from 1 to 3. In Ref.\cite{Ke:2019bkf} we set the value of
$\lambda$ within a range of $0\sim 4$ by which as believed, a bound state of two
hadrons can be formed. When the obtained $\lambda$ is much beyond
this range, one would conclude that the molecular bound state
may not exist or at least is not a stable state. But it is really
noted that the form factor is phenomenologically introduced and
the parameter $\lambda$ is usually fixed via fitting data, i.e.
neither the form factor nor the value of $\lambda$ are derived
from an underlying theory, but based on our intuition (or say, a
theoretical guess). Since the concerned processes are dominated by
the non-perturbative QCD effects whose energy scale is about 200
MeV, we have reason to believe that the cutoff should fall within
a range around a few hundreds of MeV to 1 GeV, and by this
allegation one can guess that the value of $\lambda$ should be
close to unit. However, from other aspect, this guess does not
have a solid support, further phenomenological studies and  a
better understanding on low energy field theory are needed to get
more knowledge on the form factor and value of $\lambda$. So
far, even though we believe this range for $\lambda$ which sets
a criterion to draw our conclusion at present, we cannot
absolutely rule out the possibility that some other values of
$\lambda$ beyond the designated region may hold. That is why we
proceed to compute the decay rates of $Y(4626)$ based on the
molecule postulate. (see below the numerical results for clarity
of this point).

By our strategy, for the state $Y_2$ we let $\Delta E=0.021$ GeV
which is the binding energy of the molecular state as
$M_{D^*_s}+M_{D_{s1}(2536)}-M_{Y(4626)}$. Then we try to solve the
equation $|A(\Delta E,\Lambda)-I|=0$ by varying $\lambda$ within a
reasonable range. In other words, we are trying to determine a
$\lambda$ whose value falls in the range of 0 to 4 as suggested in
literature, to satisfy the equation.

As our result, we have searched a solution of $\lambda$ within a
rather large region, but unfortunately find that there is no
solution which can satisfies the equation.

\begin{figure}[hhh]
\begin{center}
\scalebox{0.8}{\includegraphics{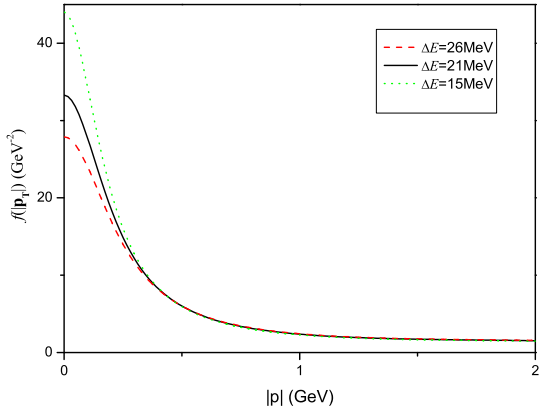}}
\end{center}
\caption{The normalized wave function $f(|\mathbf{p}_T|)$ for
$D^*_S\bar D_{s1}$ }\label{wave}
\end{figure}

However, for the $Y_1$ state if one still keeps $\Delta E=0.021$
GeV but sets $\lambda=10.59$\footnote{If the propagator of $\phi(1020)$ is
$-g_{\chi\gamma}$ in Eq. (\ref{kernel}) i.e. gauge-fixing parameter is 1 and we obtain $\lambda=10.21$ with $\Delta E=0.21$ MeV when the contributions
induced by $\eta$, $\eta'$, $f_0(980)$ and $\phi(1020)$ are included. The results indicate that
the $\phi$-exchange contribution is not very sensitive to the
choice of gauge-fixing parameter in the propagator.}, the equation $|A(\Delta
E,\lambda)-I|=0$ holds while the contributions induced by
exchanging $\eta$, $\eta'$, $f_0(980)$ and $\phi$ are included.
Instead, if the contribution of exchanging $f_0(980)$ (Fig.
\ref{t1b}) is ignored, with the same $\Delta E$ one could get a
value 10.46 of $\lambda$ which is very close to that without the
contribution of $f_0(980)$. It means that the contribution from
exchanging $f_0(980)$ is very small and can be ignored safely. On
this basis we continue to ignore the contribution from exchanging
$\phi$ and fix $\lambda=10.52$, it means that the contribution of
$\phi$ is negligible. Therefore we will only consider the
contributions from exchanging $\eta$ and $\eta'$ in latter
calculations. Meanwhile by solving the eigenequation we obtain the
wavefunction $f(Q_1),f(Q_2)...f(Q_{129})$. The normalized
wavefunction is depicted in Fig. \ref{wave} with different $\Delta
E$.

Due to existence of an error tolerance on measurements of the mass spectrum,
we are allowed to vary $\Delta E$ within a reasonable range to fix the
values of $\lambda$ again,
for the $D_{s1}\bar D^*_s$ system, the results are presented in Tab. \ref{Tab:1}.
Apparently for a reasonable
$\Delta E$ any  $\lambda$ value which is obtained by solving
the discrete B-S equation is far beyond 4.
Does the result imply that $D_{s1}\bar
D^*_s$ fails to form a bound state? We will further discuss its physical significance
in next section.

A new resonance $Y(4626)$ has been experimentally
observed\cite{Jia:2019gfe}, and it is the fact everybody acknowledges,
but what composition it has, demands a theoretical interpretation.
The molecular state explanation is favored by an intuitive observation.
However our theoretical study does not
support the allegation that $Y(4626)$ is the molecule of
$D^*_s\bar D_{s1}$.

On other respect, the above conclusion is based on a requirement:
$\lambda$ must fall in a range of 0$\sim$4, which is determined
by phenomenological studies done by many researchers. However,
$\lambda$ being in 0$\sim$4 is by no means a mandatory condition
because it is not deduced form an underlying principle and lacks
real foundation. Therefore even though our result does not favor
the molecular structure for $Y(4626)$, we still proceed to study the transitions
$Y(4626)\to D^*_{s}\bar D_{s}(2317)$, $Y(4626)\to D_{s}\bar
D_{s}(2460)$, $Y(4626)\to D_{s}(2460)\bar D^*_{s}$, $Y(4626)\to
D^*_{s}\bar D_{s}(2460)$, $Y(4626)\to D_{s}\bar D_{s2}(2573)$ and
$Y\to D_{s}\bar D_{s1}(2536)$ under the assumption of the
molecular composition of $D^*_s\bar D_{s1}$.

Using the wave function we calculate the form factors $g_0$,
$g_2$, $g'_0$, $g'_2$, $g_{10}$, $g_{11}$, $g_{12}$, $g'_{10}$,
$g'_{11}$, $g' _{12}$, $g_{20}$, $g''_0$, $g''_2$ defined in Eqs.
(\ref{a3}, \ref{b3}, \ref{c3}, \ref{d3}, \ref{e3} and \ref{f3}).
With these form factors we get the decay widths of
$Y(4626)\to D^*_{s}\bar D_{s}(2317)$, $Y(4626)\to D_{s}\bar
D_{s}(2460)$, $Y(4626)\to D_{s}(2460)\bar D^*_{s}$, $Y(4626)\to
D^*_{s}\bar D_{s}(2460)$, $Y(4626)\to D_{s}\bar D_{s1}(2573)$ and
$Y(4626)\to D_{s}\bar D_{s2}(2536)$ which are denoted as $\Gamma_a,
\Gamma_b, \Gamma_c, \Gamma_d, \Gamma_e$ and $\Gamma_f$ presented in Table
\ref{Tab:2}. Theoretical uncertainties originate from the
experimental errors, namely the theoretically predicted curve
expands to a band.

Of course, exchanging two $\eta$ ($\eta'$) mesons can also induce
a potential as the next-to-leading order (NLO) contribution, but
it undergoes a loop  suppression  therefore, we do not consider
that contribution i.e.  one-boson-exchange model is employed in
our whole scenario.

\begin{table}
\caption{the cutoff parameter $\lambda$ and the corresponding
binding energy $\Delta E$ for the bound state $D^*_s \bar D_{s1}$
}\label{Tab:1}
\begin{ruledtabular}
\begin{tabular}{cccccccc}
  $\Delta E$ (MeV)  &5 & 10 &  15  &  21& 26 \\\hline
  $\lambda$  & 10.14 & 10.28    &10.39 &10.52   &10.61
\end{tabular}
\end{ruledtabular}
\end{table}

\begin{table}
\caption{the decay widths (in units of keV) for the
transitions}\label{Tab:2}
\begin{ruledtabular}
\begin{tabular}{cccccc}
  $\Gamma_{a}$& $\Gamma_{b}$& $\Gamma_{c}$
&$\Gamma_{d}$ & $\Gamma_{e}$  & $\Gamma_{f}$\\\hline
60.6$\sim$189  & 127$\sim$342 &97.8$\sim$102
 &21.2$\sim$23.1
 &7.89$\sim$8.36&61.9$\sim$70.1
\end{tabular}
\end{ruledtabular}
\end{table}

\section{conclusion and discussion}

In this work we explore the bound state composed of a vector and
an axial vector within the B-S equation framework. Concretely we
study the resonance $Y(4626)$ which is assumed to be a molecular
state made of $D^*_s$ and $\bar D_{s1}(2536)$.  According to the
Lorentz structure we construct the B-S wave function of a vector
meson and an axial one. Using the effective interactions induced
by exchanging one light meson,  the interaction kernel is obtained
and  the B-S equation for the $D^*_s\bar D_{s1}(2536)$ system is
established. In our calculation exchanging $\eta$-meson provides
the dominant contribution (even though the contribution from
$\eta'$ is smaller than that from $\eta$, we retain it in our
calculations) while that induced by exchanging $f_0(980)$ and
$\phi(1080)$ can be safely neglected.

Under the covariant instantaneous approximation the four-dimension
B-S equation can reduce into a three-dimension B-S equation.
Integrating out the azimuthal component of the momentum we obtain
a one-dimension B-S equation which is an integral equation. With
all input parameters such as the coupling constants and the
corresponding masses of mesons we solve the equation for the
molecular state of $D^*_s\bar D_{s1}(2536)$. When we input the
binding energy $\Delta E=M_{Y(4626)}-M_{D^*_s}-M_{\bar
D_{s1}(2536)}$, we search for $\lambda$ which satisfies the
one-dimension B-S equation. Our criterion is that if there is no
solution for $\lambda$ or the value of $\lambda$ is not
reasonable, the bound state should not exist in the nature. On
contrary, if a ``suitable" $\lambda$ is found as a solution of the
B-S equation, we would claim that resonance could be a molecular
state. From the results shown in table \ref{Tab:1} one can find
that even for a small binding energy (we deliberately vary the
value of the binding energy), the $\lambda$ which makes the
equation to hold, must be larger than 9 which is far beyond the
favorable one in literature so that we tend to think the molecular
state of $D^*_s\bar D_{s1}(2536)$ does not exist unless the
coupling constants get larger than those given in Appendix.

As aforementioned discussion, the $\lambda$ in the form factor at
each vertex is phenomenologically introduced and does not receive
a solid support from any underlying principle, therefore, we may
suspect its application regime which might be the pitfall of the
phenomenology. Thus we try to overcome this barrier to extend the
value to a region which obviously deviates from the region favored
by the previous works. As a $\lambda$   value beyond 10, the
solution of the B-S equation exists, and the B-S wavefunction is
constructed.  Just using the wavefunctions, we calculate the decay
rates of $Y(4626)\to D^*_{s}\bar D_{s}(2317)$, $Y(4626)\to
D_{s}\bar D_{s}(2460)$, $Y(4626)\to D_{s}(2460)\bar D^*_{s}$,
$Y(4626)\to D^*_{s}\bar D_{s}(2460)$, $Y(4626)\to D_{s}\bar
D_{s2}(2573)$ and $Y(4626)\to D_{s}\bar D_{s2}(2536)$  under the
assumption that $Y(4626)$ is a bound state of $D^*_s\bar
D_{s1}(2536)$. Our results indicate the decay widths are small
comparing with the total width of $Y(4626)$.

The important and detectable issue is the decay patterns deduced above. This would
compose a crucial challenge to the phenomenological scenario. If the decay patterns
deduced in terms of the molecule assumption are confirmed (within an error tolerance)
, it would imply that the constraint on the phenomenological application
of form factor which is originating from the chiral perturbation can be extrapolated
to a wider region. By contrary, if the future measurements negate the predicted decay
patterns, one should claim that the assumption  that $Y(4626)$ is a molecular state
of $D^*_s\bar D_{s1}(2536)$ fails. The resonance would be in different structure, such as
tetraquark or hybrid etc.

We lay our hope on the future
experimental measurements on those decay portals, which can help us to
clarify the structure of $Y(4626)$.

\section*{Acknowledgement}
This work is supported by the National Natural Science Foundation
of China (NNSFC) under Contract No. 11375128, 11675082, 11735010
and 11975165. One of us (Hong-Wei Ke) thanks Prof. Zhi-Hui Guo for his valuable discussion.

\appendix
\section{The effective interactions}

The effective interactions can be found
in\cite{Colangelo:2005gb,Colangelo:2012xi,Ding:2008gr,Casalbuoni:1996pg,Casalbuoni:1992gi,Casalbuoni:1992dx}
\begin{eqnarray}
&&\mathcal{L}_{_{D^*D_1P}}=g_{_{D^*D_1P}}[3D^\mu_{1b}(\partial_\mu\partial_\nu
\mathcal{M})_{ba}D^{*\nu\dag}_a-D^\mu_{1b}(\partial^\nu\partial_\nu
\mathcal{M})_{ba}D^{*\dag}_{a\mu}+\frac{1}{m_{D^*}m_{D_1}}\partial^\nu
D^\mu_{1b}(\partial_\nu\partial_\tau
\mathcal{M})_{ba}\partial^\tau
D^{*\dag}_{a\mu}]\nonumber\\&&+g_{_{\bar D^*\bar D_1P}}[3\bar
D^\mu_{1b}(\partial_\mu\partial_\nu \mathcal{M})_{ba}\bar
D^{*\nu\dag}_a-\bar D^\mu_{1b}(\partial^\nu\partial_\nu
\mathcal{M})_{ba}\bar
D^{*\dag}_{a\mu}+\frac{1}{m_{D^*}m_{D_1}}\partial^\nu \bar
D^\mu_{1b}(\partial_\nu\partial_\tau
\mathcal{M})_{ba}\partial^\tau \bar
D^{*\dag}_{a\mu}]\nonumber\\&&+c.c.,\\&&\mathcal{L}_{_{D_0D_1P}}=g_{_{D_0D_1P}}D^\mu_{1b}(\partial_\mu
\mathcal{M})_{ba}D^{\dag}_{0a}+g_{_{\bar D_0\bar D_1P}}\bar
D^\mu_{1b}(\partial_\mu \mathcal{M})_{ba}\bar
D^{\dag}_{0a}+c.c.,\\&&\mathcal{L}_{_{D^*D^*P}}=g_{_{D^*D^*P}}(D^{*\mu}_{b}\stackrel{\leftrightarrow}{\partial}^{\beta}
D^{*\alpha\dag}_{a})(\partial^\nu
\mathcal{M})_{ba}\varepsilon_{\nu\mu\alpha\beta}+\nonumber\\&&g_{_{\bar
D^*\bar D^*P}}(\bar
D^{*\mu}_{b}\stackrel{\leftrightarrow}{\partial}^{\beta} \bar
D^{*\alpha\dag}_{a})(\partial^\nu
\mathcal{M})_{ba}\varepsilon_{\nu\mu\alpha\beta}+c.c.,\\&&\mathcal{L}_{_{D_{1}D_{1}P}}=g_{_{D_1D_1P}}(D^{\mu}_{1b}\stackrel{\leftrightarrow}{\partial}^{\beta}
D^{\alpha\dag}_{1a})(\partial^\nu
\mathcal{M})_{ba}\varepsilon_{\mu\nu\alpha\beta}+\nonumber\\&&g_{_{\bar
D_1\bar D_1P}}(\bar
D^{\mu}_{1b}\stackrel{\leftrightarrow}{\partial}^{\beta} \bar
D^{\alpha\dag}_{1a})(\partial^\nu
\mathcal{M})_{ba}\varepsilon_{\mu\nu\alpha\beta}+c.c.,\\&&\mathcal{L}_{_{DD^*P}}=g_{_{DD^*P}}D_{b}(\partial_\mu
\mathcal{M})_{ba}D^{*\mu\dag}_{a}+g_{_{DD^*P}}D^{*\mu}_{b}(\partial_\mu
\mathcal{M})_{ba}D^{\dag}_{a}+\nonumber\\&&g_{_{\bar D\bar
D^*P}}\bar D_{b}(\partial_\mu \mathcal{M})_{ba}\bar
D^{*\mu\dag}_{a}+g_{_{\bar D\bar D^*P}}\bar
D^{*\mu}_{b}(\partial_\mu \mathcal{M})_{ba}\bar
D^{\dag}_{a}+c.c.,\\&&\mathcal{L}_{_{D^*D'_1P}}=ig_{_{D^*D'_1P}}[
\frac{\partial^\alpha D^{*\mu}_{b}(\partial_\alpha
\mathcal{M})_{ba}D'^\dag_{1a\nu}}{M_{D_1}}-
\frac{D^{*\mu}_{b}(\partial_\alpha
\mathcal{M})_{ba}\partial^\alpha
D'^\dag_{1a\nu}}{M_{D^*}}]+\nonumber\\&&ig_{_{\bar D^*\bar
D'_1P}}[ \frac{\partial^\alpha \bar D^{*\mu}_{b}(\partial_\alpha
\mathcal{M})_{ba}\bar D'^\dag_{1a\nu}}{M_{D_1}}- \frac{\bar
D^{*\mu}_{b}(\partial_\alpha \mathcal{M})_{ba}\partial^\alpha \bar
D'^\dag_{1a\nu}}{M_{D^*}}]+c.c.,\\&&\mathcal{L}_{_{D_{1}D'_{1}P}}=g_{_{D_1D'_1P}}(\frac{{\partial}^{\beta}D^{\mu}_{1b}
D^{\alpha\dag}_{1a}}{m_{D_1}}-\frac{D^{\mu}_{1b}
{\partial}^{\beta}D^{\alpha\dag}_{1a}}{m_{D'_1}})(\partial^\nu
\mathcal{M})_{ba}\varepsilon_{\mu\nu\alpha\beta}+\nonumber\\&&g_{_{\bar
D_1\bar D'_1P}}(\frac{{\partial}^{\beta}\bar D^{\mu}_{1b} \bar
D^{\alpha\dag}_{1a}}{m_{D_1}}-\frac{\bar D^{\mu}_{1b}
{\partial}^{\beta}\bar
D^{\alpha\dag}_{1a}}{m_{D'_1}})(\partial^\nu
\mathcal{M})_{ba}\varepsilon_{\mu\nu\alpha\beta}+c.c.,\\&&\mathcal{L}_{_{D_{1}D_{2}P}}=g_{_{D_1D_2P}}(D_{1a\mu})(\partial_\nu\mathcal{M})_{ba}D_{2a}^{\dag\mu\nu}
+g_{_{\bar D_1\bar D_2P}}(\bar
D_{1a\mu})(\partial_\nu\mathcal{M})_{ba}\bar
D_{2a}^{\dag\mu\nu}+c.c.,
\\&&\mathcal{L}_{_{D_{1}D_{1}f_0}}=g_{_{D_1D_1f_0}}(D^\mu_{1a})D{\dag}_{1a\mu}f_0+g_{_{\bar D_1\bar D_1f_0}}(\bar D^\mu_{1a})\bar D{\dag}_{1a\mu}f_0+c.c.,
\\&&\mathcal{L}_{_{D^*D^*f_0}}=g_{_{D^*D^*f_0}}(D^{*\mu}_{a})D^{*\dag}_{a\mu}f_0+g_{_{\bar D^*\bar D^*f_0}}(\bar D^{*\mu}_{a})\bar
D^{*\dag}_{a\mu}f_0+c.c.,
\\&&\mathcal{L}_{_{D_{1}D^*f_0}}=ig_{D_{1}D^*f_0}\varepsilon_{\mu\alpha\nu\beta}
(D^{\mu}_{1a}\stackrel{\leftrightarrow}{\partial}^{\alpha}
D^{*\nu\dag}_{a}\partial^\beta
f_0+D^{*\mu\dag}_{a}\stackrel{\leftrightarrow}{\partial}^{\alpha}
D^{\nu}_{1a}\partial^\beta f_0+\bar D^{\mu}_{b}
\stackrel{\leftrightarrow}{\partial}^{\alpha}\bar
D^{*\nu\dag}_{a}\partial^\beta f_0\nonumber\\&&+\bar
D^{*\mu\dag}_{b}\stackrel{\leftrightarrow}{\partial}^{\alpha} \bar
D^{\nu}_{a}\partial^\beta f_0)+c.c.,
\\&&\mathcal{L}_{_{D_{1}D_{1}V}}=ig_{_{D_1D_1V}}(D^{\nu}_{1b}\stackrel{\leftrightarrow}{\partial}_{\mu}
D^{\dag}_{1a\nu})(
\mathcal{V})_{ba}^\mu+ig'_{_{D_1D_1V}}(D^{\mu}_{1b}
D^{\nu\dag}_{1a}-D^{\mu\dag}_{1b} D^{\nu}_{1a})(
\partial_\mu\mathcal{V}_\nu-\partial_\nu\mathcal{V}_\mu)_{ba}\nonumber\\&&+ig_{_{\bar D_1\bar D_1V}}(\bar D^{\nu}_{1b}\stackrel{\leftrightarrow}{\partial}_{\mu}
\bar D^{\dag}_{1a\nu})( \mathcal{V})_{ba}^\mu+ig'_{_{\bar D_1\bar
D_1V}}(\bar D^{\mu}_{1b} \bar D^{\nu\dag}_{1a}-\bar
D^{\mu\dag}_{1b} \bar D^{\nu}_{1a})(
\partial_\mu\mathcal{V}_\nu-\partial_\nu\mathcal{V}_\mu)_{ba}+c.c.,
\\&&\mathcal{L}_{_{D^*D^*V}}=ig_{_{D^*D^*V}}(D^{*\nu}_{b}\stackrel{\leftrightarrow}{\partial}_{\mu}
D^{*\dag}_{a\nu})(
\mathcal{V})_{ba}^\mu+ig'_{_{D^*D^*V}}(D^{*\mu}_{b}
D^{*\nu\dag}_{a}-D^{*\mu\dag}_{b} D^{*\nu}_{a})(
\partial_\mu\mathcal{V}_\nu-\partial_\nu\mathcal{V}_\mu)_{ba}\nonumber\\&&+ig_{_{\bar D^*\bar D^*V}}(\bar D^{*\nu}_{b}\stackrel{\leftrightarrow}{\partial}_{\mu}
\bar D^{*\dag}_{a\nu})( \mathcal{V})_{ba}^\mu+ig'_{_{\bar D^*\bar
D^*V}}(\bar D^{*\mu}_{b} \bar D^{*\nu\dag}_{a}-\bar
D^{*\mu\dag}_{b} \bar D^{*\nu}_{a})(
\partial_\mu\mathcal{V}_\nu-\partial_\nu\mathcal{V}_\mu)_{ba}+c.c.\\&&\mathcal{L}_{_{D_{1}D^*V}}=ig_{D_{1}D^*V}\varepsilon_{\mu\nu\alpha\beta}
(D^{\mu}_{1b}\stackrel{\leftrightarrow}{\partial}^{\alpha}
D^{*\nu\dag}_{a}+D^{*\mu\dag}_{b}\stackrel{\leftrightarrow}{\partial}^{\alpha}
D^{\nu}_{1a}+\bar D^{\mu}_{1b}
\stackrel{\leftrightarrow}{\partial}^{\alpha}\bar
D^{*\nu\dag}_{a}+\bar
D^{*\mu\dag}_{b}\stackrel{\leftrightarrow}{\partial}^{\alpha} \bar
D^{\nu}_{1a})(
\mathcal{V}^\beta)_{ba}\nonumber\\&&+g'_{D_{1}D^*V}\varepsilon_{\mu\nu\alpha\beta}
(D^{\mu}_{1b} D^{*\nu\dag}_{a}+D^{*\mu\dag}_{b} D^{\nu}_{1a}+\bar
D^{\mu}_{1b} \bar D^{*\nu\dag}_{a}+\bar D^{*\mu\dag}_{b} \bar
D^{\nu}_{1a})(
\partial^\alpha\mathcal{V}^\beta)_{ba}+c.c.,
\end{eqnarray}
where $c.c.$ is the complex conjugate term, $a$ and $b$ represent
the flavors of light quarks, $f_0$ denotes $f_0(980)$. In
Refs.\cite{Ding:2008gr} $\mathcal{M}$ and $\mathcal{V}$ are
$3\times 3$ hermitian and traceless matrixs $
\left(\begin{array}{ccc}
        \frac{\pi^0}{\sqrt{2}}+\frac{\eta}{\sqrt{6}} &\pi^+ &K^+ \\
         \pi^- & -\frac{\pi^0}{\sqrt{2}}+\frac{\eta}{\sqrt{6}}&K^0\\
         K^-& \bar{K^0} & -\sqrt{\frac{2}{3}}\eta
      \end{array}\right)$
       and $
\left(\begin{array}{ccc}
        \frac{\rho^0}{\sqrt{2}}+\frac{\omega}{\sqrt{2}} &\rho^+ &K^{*+} \\
         \rho^- & \frac{\rho^0}{\sqrt{2}}+\frac{\omega}{\sqrt{2}}&K^{*0}\\
         K^{*-}& \bar{K^{*0}} & \phi
      \end{array}\right)$ respectively. Now in order to study the coupling of $\eta'$ with $D^*_S$ and $D_{s1}$ following Ref.\cite{Guo:2015xva} we need extend $\mathcal{M}$ to $ \left(\begin{array}{ccc}
        \frac{\pi^0}{\sqrt{2}}+\frac{\eta_8}{\sqrt{6}}+\frac{\eta_0}{\sqrt{3}} &\pi^+ &K^+ \\
         \pi^- & -\frac{\pi^0}{\sqrt{2}}+\frac{\eta_8}{\sqrt{6}}+\frac{\eta_0}{\sqrt{3}}&K^0\\
         K^-& \bar{K^0} & -\sqrt{\frac{2}{3}}\eta_8+\frac{\eta_0}{\sqrt{3}}
      \end{array}\right)$  where $\eta_8$ and $\eta_0$ are $SU(3)$ octet and singlet. The physical states $\eta$ and $\eta'$ are the mixtures of $\eta_8$ and $\eta_0$:  $\eta={\rm cos\theta}\eta_8-{\rm sin\theta}\eta_0$
and
      $\eta'={\rm sin\theta}\eta_8+{\rm cos\theta}\eta_0$. In order to keep the derived interactions involving $\eta$ unchanged compared
with those formulaes given in reference[37-39] we set the mixing angle $\theta$ to 0 so
$ \mathcal{M}=\left(\begin{array}{ccc}
        \frac{\pi^0}{\sqrt{2}}+\frac{\eta}{\sqrt{6}}+\frac{\eta'}{\sqrt{3}} &\pi^+ &K^+ \\
         \pi^- & -\frac{\pi^0}{\sqrt{2}}+\frac{\eta}{\sqrt{6}}+\frac{\eta'}{\sqrt{3}}&K^0\\
         K^-& \bar{K^0} & -\sqrt{\frac{2}{3}}\eta+\frac{\eta'}{\sqrt{3}}
      \end{array}\right)$. In Ref.\cite{Guo:2015xva} the authors estimated $\theta$ and obtained it as $-18.9^\circ$
so the approximation holds roughly.

In the chiral and heavy quark limit, the above coupling constants
are
$$g_{_{D^*_sD_{s1}\eta}}=g_{_{\bar D^*_s\bar D_{s1}\eta}}=-\sqrt{2}g_{_{D^*_sD_{s1}\eta'}}=-\sqrt{2}g_{_{\bar D^*_s\bar D_{s1}\eta'}}=-\frac{\sqrt{6}}{3}\frac{h_1+h_2}{\Lambda_{\chi}f_{\pi}}\sqrt{M_{D^*_{s}}M_{D_{s1}}},$$
$$g_{_{D_{s0}D_{s1}\eta}}=g_{_{\bar D_{s0}\bar D_{s1}\eta}}=-\sqrt{2}g_{_{D_{s0}D_{s1}\eta'}}=-\sqrt{2}g_{_{\bar D_{s0}\bar D_{s1}\eta'}}=-\frac{2\sqrt{6}}{3}\frac{\tilde{h}}{f_{\pi}}\sqrt{M_{D_{s0}}M_{D_{s1}}},$$
$$g_{_{D^*_sD^*_s\eta}}=g_{_{\bar D^*_s\bar D^*_s\eta}}=-\sqrt{2}g_{_{D^*_sD^*_s\eta'}}=-\sqrt{2}g_{_{\bar D^*_s\bar D^*_s\eta'}}=\frac{g}{f_\pi},$$
$$g_{_{D_{s1}D_{s1}\eta}}=g_{_{\bar D_{s1}\bar D_{s1}\eta}}=-\sqrt{2}g_{_{D_{s1}D_{s1}\eta'}}=-\sqrt{2}g_{_{\bar D_{s1}\bar D_{s1}\eta'}}=\frac{5\kappa}{6f_\pi},$$
$$g_{_{D_sD^*_s\eta}}=-g_{_{\bar D_s\bar D^*_s\eta}}=-\sqrt{2}g_{_{D_sD^*_s\eta'}}=\sqrt{2}g_{_{\bar D_s\bar D^*_s\eta'}}=-\frac{2g}{f_{\pi}}
\sqrt{M_{D_s}M_{D_s^*}},$$
$$g_{_{D^*_sD'_{s1}\eta}}=g_{_{\bar D^*_s\bar D'_{s1}\eta}}=-\sqrt{2}g_{_{D^*_sD'_{s1}\eta'}}=-\sqrt{2}g_{_{\bar D^*_s\bar D'_{s1}\eta'}}=\frac{h}{f_{\pi}}
\sqrt{M_{D^*_s}M_{D'_{s1}}},$$
$$g_{_{D_{s1}D'_{s1}\eta}}=g_{_{\bar D_{s1}\bar D'_{s1}\eta}}=-\sqrt{2}g_{_{D_{s1}D'_{s1}\eta'}}=-\sqrt{2}g_{_{\bar D_{s1}\bar D'_{s1}\eta'}}=\frac{\sqrt{6}\tilde{h}}{6f_{\pi}}
\sqrt{M_{D_{s1}}M_{D'_{s1}}},$$
$$g_{_{D_{s1}D_{s2}\eta}}=g_{_{\bar D_{s1}\bar D_{s2}\eta}}=-\sqrt{2}g_{_{D_{s1}D_{s2}\eta'}}=-\sqrt{2}g_{_{\bar D_{s1}\bar D_{s2}\eta'}}=-\frac{\sqrt{6}\kappa}{3f_{\pi}}
\sqrt{M_{D_{s1}}M_{D_{s2}}},$$
$$g_{_{D^*_sD^*_s\phi}}=-g_{_{\bar D^*_s\bar D^*_s\phi}}=-\frac{\beta g_V}{\sqrt{2}},\,\,
g'_{_{D^*_sD^*_s\phi}}=-g'_{_{\bar D^*_s\bar
D^*_s\phi}}=-\sqrt{2}\lambda g_V M_{D^*_s}$$
$$g_{_{D_{s1}D_{s1}\phi}}=g_{_{\bar D_{s1}\bar D_{s1}\phi}}=\frac{\beta_2 g_V}{\sqrt{2}},\,\,
g'_{_{D_{s1}D_{s1}\phi}}=g'_{_{\bar D_{s1}\bar
D_{s1}\phi}}=\frac{5\lambda_2 g_V}{3\sqrt{2}}M_{D_{s1}},$$
$$g_{_{D^*_sD_{s1}\phi}}=g_{_{\bar D^*_s\bar D_{s1}\phi}}=\frac{g_V\zeta_1}{2\sqrt{3}},\,\,
g_{_{D^*_sD_{s1}\phi}}=g_{_{\bar D^*_s\bar
D_{s1}\phi}}=\frac{2g_V\mu_1}{2\sqrt{3}}$$
 and we suppose
$$g_{_{D_{s}^*D_{s}^*f_0}}=g_{_{D*D*\sigma}}=-2g_{\sigma}M_{D_{s}^*},$$
$$g_{_{D_{s1}D_{s1}f_0}}=g_{_{D_{1}D_{1}\sigma}}=-2g''_{\sigma}M_{D_{s1}},$$
$$g_{_{D_{s1}D^*_{s}f_0}}=g_{_{D_{1}D^*\sigma}}=i\frac{h'_\sigma}{\sqrt{6}f_\pi}.$$

with $\Lambda_{\chi}=1$GeV, $f_\pi=132$
MeV\cite{Colangelo:2005gb}, $h=0.56$, $h_1=h_2=0.43$,
$g=0.64$\cite{Colangelo:2012xi}, $\kappa=g$,
$\tilde{h}=0.87$\cite{Falk:1992cx},$g_{\sigma}=0.761$\cite{Bardeen:2003kt},
$g''_{\sigma}=g_{\sigma}$, $h'_\sigma=0.346$\cite{Liu:2008xz},
$\beta=0.9$, $g_V=5.9$, $\lambda_1=0.56$\cite{Falk:1992cx},
$\beta_2=1.1$, $\lambda_2=-0.6$ $\zeta_1=-0.1$\cite{He:2019csk},
$\mu_1=0$\cite{Dong:2019ofp}.

\end{document}